\documentclass[10pt,letterpaper]{article}
\usepackage{blindtext,titlefoot}
\usepackage{authblk}
\usepackage{marvosym}
\usepackage[utf8]{inputenc}
\usepackage[english]{babel}
\usepackage{amsmath}
\usepackage{amsfonts}
\usepackage{amssymb}
\usepackage{makeidx}
\usepackage{graphicx}
\usepackage{lmodern}
\usepackage{kpfonts}
\usepackage{dsfont}
\usepackage{cancel}
\usepackage{amsthm} 
\usepackage{subcaption}
\usepackage{multirow}
\usepackage{float} 
\usepackage{url}
\usepackage{breakurl}
\usepackage[breaklinks]{hyperref}
\usepackage{multicol}
\usepackage{mathtools, cuted}
\usepackage{lipsum}
\usepackage{booktabs}
\usepackage{comment}
\usepackage{cite}
\usepackage{longtable}
\usepackage{listings}

\newtheorem{theorem}{ Theorem}[section]

\newtheorem{corollary}[theorem]{Corollary}
\newtheorem{example}{Example}
\usepackage[utf8]{inputenc}
\usepackage[english]{babel}
\usepackage{amsmath}
\usepackage{amsfonts}
\usepackage{amssymb}
\usepackage{makeidx}
\usepackage{graphicx}
\usepackage{lmodern}
\usepackage{kpfonts}
\usepackage[left=1.3cm,right=1.3cm,top=2cm,bottom=2cm]{geometry}
\usepackage{dsfont}
\usepackage{hyperref}
\usepackage{verbatim}
\usepackage{cite} 
\usepackage{caption}
\usepackage{subcaption}
\usepackage{fancyhdr}
\DeclareMathOperator{\Op}{O}
\DeclareMathOperator{\MOp}{MO}

\DeclareMathOperator{\Og}{G}

\DeclareMathOperator{\Ops}{S}
\DeclareMathOperator{\Opt}{T}

\newcommand{\nset}[1]{
\mathbb{#1}
}
\newcommand{\set}[1]{
\left\{#1\right\}
}
\newcommand{\com}[1]{``#1''}

\newcommand{\ifr}[5]{
{}^{#1}_{#2}{#3}_{#4}^{#5}
}

\newcommand{\gam}[1]{
\Gamma\left(#1 \right)
}

\newcommand{\norm}[1]{
\left\lVert #1 \right\rVert
}
\newcommand{\abs}[1]{
\left\lvert #1 \right\rvert
}
\newcommand{\ds}[1]{
\displaystyle #1
}
\newcommand{\der}[3]{
\dfrac{#1^{#3} }{ #1 #2^{#3}}
}

\newcommand{\mspc}[2]{
\begin{array}{c}
\ds #1 \vspace{#2}
\end{array}
}

\usepackage{enumitem} 
\setlist[itemize]{noitemsep} 

\usepackage{titlesec} 
\titleformat{\section}[block]{\large\bfseries\scshape\centering}{\thesection.}{1em}{} 
\titleformat{\subsection}[block]{\large\bfseries\scshape\centering}{\thesubsection.}{1em}{}
\titleformat{\subsubsection}[block]{\large\bfseries\scshape\centering}{\thesubsubsection.}{1em}{} 

\title{\huge \bfseries Proposal for the Application of Fractional Operators in Polynomial Regression Models to Enhance the Determination Coefficient $R^2$ on Unseen Data}

\author[,a,b]{Anthony Torres-Hernandez  \footnote{Email: anthony.torres@ciencias.unam.mx; Corresponding author; ORCID: 0000-0001-6496-9505}}

\affil[a]{\normalsize Department of Information and Communication Technologies, Music and Machine Learning Lab, Universitat Pompeu Fabra, Barcelona}

\affil[b]{\normalsize Department of Physics, Faculty of Science, Universidad Nacional Autónoma de México, Mexico City}

\date{}

\begin{document}

\maketitle


\begin{abstract}

Since polynomial regression models are generally quite reliable for data with a linear trend, it is important to note that, in some cases, they may encounter overfitting issues during the training phase, which could result in negative values of the coefficient of determination $R^2$ for unseen data. For this reason, this work proposes the partial implementation of fractional operators in polynomial regression models to generate a fractional regression model. The goal of this proposal is to attempt to mitigate overfitting, which could improve the value of the coefficient of determination for unseen data, compared to the polynomial model, under the assumption that this would contribute to generating predictive models with better performance. The methodology for constructing these fractional regression models is detailed, and examples applicable to both Riemann-Liouville and Caputo fractional operators are presented.

\end{abstract}

\section{Introduction}

Fractional calculus is a branch of mathematics that uses derivatives of non-integer order that originated around the same time as conventional calculus due to Leibniz's notation for derivatives of integer~order
\begin{eqnarray*}
\dfrac{d^n}{d x^n}.
\end{eqnarray*}

Thanks to this notation, L'Hopital could ask in a letter to Leibniz about the interpretation of taking $ n = 1/2 $ in a derivative. Since at that moment Leibniz could not give a physical or geometrical interpretation of this question, he simply answered to L'Hopital in a letter, \com{$\dots$ is an apparent paradox of which, one day, useful consequences will be drawn}~\cite{miller93}. The~name of fractional calculus comes from a historical question since, in this branch of mathematical analysis, the derivatives and integrals of a certain order $\alpha$ are studied, with~$\alpha \in \nset{R}$. Currently, fractional calculus does not have a unified definition of what is considered a fractional derivative. As~a consequence, when it is not necessary to explicitly specify the form of a fractional derivative, it is usually denoted as~follows

\begin{eqnarray*}
\dfrac{d^\alpha}{d x^\alpha}.
\end{eqnarray*}

The fractional operators have many representations, but~one of their fundamental properties is that they allow retrieving the results of conventional calculus when $\alpha \to n$. For~example, let $f:\Omega \subset \nset{R}\to \nset{R}$ be a function such that  $f\in L_{loc}^1(a,b)$, where $L_{loc}^1(a,b)$ denotes the space of locally integrable functions on the open interval $(a,b)\subset \Omega$. One of the fundamental operators of fractional calculus is the operator {Riemann--Liouville fractional integral}
,  which is defined   as follows~\cite{hilfer00,oldham74}:
\begin{eqnarray}
\ifr{}{a}{I}{x}{\alpha}f(x):=\dfrac{1}{\gam{\alpha}}\int_a^x (x-t)^{\alpha-1}f(t)dt,
\end{eqnarray}
where $\Gamma$ denotes the Gamma function. It is worth mentioning that the above operator is a fundamental piece to construct the operator {Riemann-Liouville fractional derivative},  which is defined as follows~\cite{hilfer00,kilbas2006theory}:

\begin{eqnarray}\label{eq:1-001}
\begin{array}{c}
\ifr{}{a}{D}{x}{\alpha}f(x) := \left\{
\begin{array}{cc}
\ds \ifr{}{a}{I}{x}{-\alpha}f(x), &\mbox{if }\alpha<0 \vspace{0.1cm}\\  
\ds \dfrac{d^n}{dx^n}\left( \ifr{}{a}{I}{x}{n-\alpha}f(x)\right), & \mbox{if }\alpha\geq 0
\end{array}
\right.
\end{array}, 
\end{eqnarray}
where  $ n = \lceil \alpha \rceil$ and $\ifr{}{a}{I}{x}{0}f(x):=f(x)$. On~the other hand, let $f:\Omega \subset \nset{R}\to \nset{R}$ be a function $n$-times differentiable such that $f,f^{(n)}\in L_{loc}^1(a,b)$. Then, the~Riemann--Liouville fractional integral also allows constructing the operator {Caputo fractional derivative}, which is defined as follows~\cite{hilfer00,kilbas2006theory}:

\begin{eqnarray}\label{eq:1-002}
\begin{array}{c}
\ifr{C}{a}{D}{x}{\alpha}f(x) := \left\{
\begin{array}{cc}
\ds \ifr{}{a}{I}{x}{-\alpha}f(x), &\mbox{if }\alpha<0 \vspace{0.1cm}\\  
\ds  \ifr{}{a}{I}{x}{n-\alpha}f^{(n)}(x), & \mbox{if }\alpha\geq 0
\end{array}
\right.
\end{array}, 
\end{eqnarray}
where  $ n = \lceil \alpha \rceil$ and $\ifr{}{a}{I}{x}{0}f^{(n)}(x):=f^{(n)}(x)$. Furthermore, if~the function $f$ fulfills that $f^{(k)}(a)=0 \ \forall k\in \set{0,1,\cdots,n-1}$, the~Riemann--Liouville fractional derivative coincides with the Caputo fractional derivative, that is,

\begin{eqnarray}\label{eq:1-003}
\ifr{}{a}{D}{x}{\alpha}f(x)=\ifr{C}{a}{D}{x}{\alpha}f(x).
\end{eqnarray}

Therefore,  applying the operator \eqref{eq:1-001} with $a=0$ to the  function $ x^{\mu} $, with~$\mu> -1$, we obtain the following result:
\begin{eqnarray}\label{eq:3-002}
\ifr{}{0}{D}{x}{\alpha}x^\mu = 
 \dfrac{\gam{\mu+1}}{\gam{\mu-\alpha+1}}x^{\mu-\alpha}, & \alpha\in \nset{R}\setminus \nset{Z},
\end{eqnarray}
where if $1\leq \lceil \alpha \rceil \leq \mu$, it is fulfilled that $\ifr{}{0}{D}{x}{\alpha}x^\mu=\ifr{C}{0}{D}{x}{\alpha}x^\mu$. To illustrate a bit the diversity of representations that fractional operators may have, we proceed to present a recapitulation of some fractional derivatives, fractional integrals, and local fractional operators that may be found in the literature \cite{de2014review,teodoro2019review,valerio2022many}:

\begin{table}[h]
    \centering
    \begin{tabular}{|c|c|}
        \hline
        \textbf{Name} & \textbf{Expression} \\
        \hline
        Gr\"unwald-Letnikov fractional derivative & $\displaystyle \ifr{GL}{a}{D}{x}{\alpha}f(x)=\lim_{h\to 0}\dfrac{1}{h^\alpha}\sum_{k=0}^{n }\dfrac{(-1)^k\gam{\alpha+1}}{\gam{k+1}\gam{\alpha-k+1}}f(x-kh), \quad n= \lfloor (x-a)/h \rfloor$ \\
        \hline
        Marchaud fractional derivative & $\displaystyle \ifr{Ma}{-\infty}{D}{x}{\alpha}f(x)=\dfrac{\alpha}{\gam{1-\alpha}}\int_{-\infty}^x (x-t)^{-\alpha-1} \left( f(x)-f(t)\right) dt , \quad 0<\alpha<1$ \\
        \hline
        Hadamard fractional derivative & $\displaystyle \ifr{Ha}{a}{D}{x}{\alpha}f(x)=\dfrac{x}{\gam{1-\alpha}}\dfrac{d}{dx}\int_a^x\left(\ln(x)-\ln(t) \right)^{2-\alpha}\dfrac{f(t)}{t}dt, \quad 0<\alpha<1$ \\
        \hline
        Chen fractional derivative & $\displaystyle \ifr{Ch}{a}{D}{x}{\alpha}f(x)=\dfrac{1}{\gam{1-\alpha}}\dfrac{d}{dx}\int_a^x(x-t)^{-\alpha}f(t)dt, \quad	 0<\alpha<1.$ \\
        \hline
        Caputo-Fabrizio fractional derivative & $ \displaystyle \ifr{CF}{a}{D}{x}{\alpha}f(x)=\dfrac{M(\alpha)}{1-\alpha}\int_a^x \exp\left( -\frac{\alpha}{1-\alpha}(x-t) \right) f^{(1)}(t)dt,  \quad M(0)=M(1)=1$ \\
        \hline
        Atangana-Baleanu-Caputo fractional derivative & $\displaystyle \ifr{ABC}{a}{D}{x}{\alpha}f(x)=\dfrac{M(\alpha)}{1-\alpha}\int_a^x E_\alpha \left( -\frac{\alpha}{1-\alpha}(x-t)^\alpha \right) f^{(1)}(t)dt, \quad M(0)=M(1)=1$ \\
        \hline
        Canavati fractional derivative & $\displaystyle \ifr{Ca}{a}{D}{x}{\alpha}f(x)=\dfrac{1}{\gam{1+\alpha-n}}\dfrac{d}{dx}\int_a^x(x-t)^{n-\alpha} \der{d}{t}{n}f(t)dt, \quad n=\lfloor \alpha \rfloor$ \\
        \hline
        Jumarie fractional derivative & $\displaystyle \ifr{Ju}{a}{D}{x}{\alpha}f(x)=\dfrac{1}{\gam{n-\alpha}}\der{d}{x}{n}\int_a^x(x-t)^{n-\alpha-1}\left(f(t)-f(a) \right)dt, \quad n=\lceil \alpha \rceil$ \\
        \hline
        Hadamard fractional integral & $\displaystyle \ifr{Ha}{a}{I}{x}{\alpha}f(x)=\dfrac{1}{\gam{\alpha}}\int_a^x\left(\ln(t)-\ln(x) \right)^{\alpha-1}\dfrac{f(t)}{t}dt$ \\
        \hline
        Weyl fractional integral & $\displaystyle \ifr{}{x}{W}{\infty}{\alpha}f(x)=\dfrac{1}{\gam{\alpha}}\int_x^\infty(t-x)^{\alpha-1}f(t)dt$ \\
        \hline
        Conformable fractional operator & $\displaystyle T_\alpha f(x)=\lim_{h \to 0}\dfrac{f\left(x+h x^{1-\alpha} \right)-f(x)  }{h}$ \\
        \hline
        Katugampola fractional operator & $\displaystyle D^\alpha f(x)=\lim_{h\to 0}\dfrac{f\left(x \exp\left( hx^{-\alpha}\right) \right)-f(x) }{h}$ \\
        \hline
        Deformable fractional operator & $\displaystyle \mathcal{D}^\alpha f(x)=\lim_{h\to 0}\dfrac{(1+h\beta)f(x+h\alpha)-f(x)}{h}, \quad \alpha+\beta=1$ \\
        \hline
    \end{tabular}
    \caption{Different fractional operators}
    \label{tab:frac_operators}
\end{table}

Before~continuing, it is worth mentioning that the applications of fractional operators have spread to different fields of science, such as finance~\cite{safdari2015radial,torres2021blackscholes}, economics~\cite{traore2020model,tejado2019fractional}, number theory through the Riemann zeta function~\cite{guariglia2021fractional,torres2021zeta}, in~engineering with the study for the manufacture of hybrid solar receivers~\cite{de2021fractional,torres2020reduction}, and in physics and mathematics to solve nonlinear algebraic equation systems~\cite{erfanifar2020modified,cordero2019variant,gdawiec2021newton,gdawiec2019visual,akgul2019fractional,torres2021fracnewrap,torres2021fracnewrapaitken,torres2020fracsome,candelario2020multipoint,candelario2022optimal}, which is a classical problem in mathematics, physics and engineering that consists of finding the set of zeros of a function $f:\Omega \subset \nset{R}^n \to \nset{R}^n$, that is,
\begin{eqnarray*}
\set{\xi \in \Omega \ : \ \norm{f(\xi)}=0},
\end{eqnarray*}
where $\norm{ \ \cdot \ }: \nset{R}^n \to \nset{R}$ denotes any vector norm, or~equivalently,
\begin{eqnarray*}
\set{\xi \in \Omega \ : \ [f]_k(\xi)=0 \hspace{0.15cm} \forall k\geq 1},
\end{eqnarray*}
where $[f]_k: \nset{R}^n \to \nset{R}$ denotes the $k$-th component of the function $f$.

\section{Sets of Fractional Operators}

Before proceeding, it is important to note that the large number of fractional operators in the literature~\cite{de2014review,teodoro2019review,valerio2022many,osler1970leibniz,almeida2017caputo,fu2021continuous,fan2022note,abu2021generalized,saad2020new,rahmat2019new,sousa2018psi,jarad2017new,atangana2017new,yavuz2020comparing,liu2020new,yang2017new,atangana2016new,he2016new,sene2020fractional} suggests that the most natural way to characterize the elements of fractional calculus is through sets. This approach is central to the methodology of \textit{fractional calculus of sets} \cite{torres2021sets,torres2022acceleration}. 

Consider a scalar function $h: \mathbb{R}^m \to \mathbb{R}$ and the canonical basis of $\mathbb{R}^m$, denoted by $\{\hat{e}_k\}_{k \geq 1}$. Using Einstein’s summation convention, we can define the fractional operator of order $\alpha$ as:

\begin{equation}
o_x^\alpha h(x) := \hat{e}_k o_k^\alpha h(x).
\end{equation}

Next, denote by $\partial_k^n$ the partial derivative of order $n$ with respect to the $k$-th component of the vector $x$. Using the previous operator, we define the following set of fractional operators:

\begin{equation}\label{eq:2-001}
\Op_{x,\alpha}^n(h) := \left\{ o_x^\alpha \ : \ \exists o_k^\alpha h(x) \ \text{and} \ \lim_{\alpha \to n} o_k^\alpha h(x) = \partial_k^n h(x) \ \forall k \geq 1 \right\},
\end{equation}
which is non-empty since it includes the following set of fractional operators:

\begin{equation}
\Op_{0,x,\alpha}^n(h) := \left\{ o_x^\alpha \ : \ \exists o_k^\alpha h(x) = \left(\partial_k^n + \mu(\alpha) \partial_k^\alpha \right) h(x), \ \lim_{\alpha \to n} \mu(\alpha) \partial_k^\alpha h(x) = 0 \ \forall k \geq 1 \right\}.
\end{equation}

Consequently, the following result holds:

\begin{equation}
\text{If } o_{i,x}^\alpha, o_{j,x}^\alpha \in \Op_{x,\alpha}^n(h) \ \text{with} \ i \neq j, \ \Rightarrow \ \exists o_{k,x}^\alpha = \frac{1}{2} \left(o_{i,x}^\alpha + o_{j,x}^\alpha \right) \in \Op_{x,\alpha}^n(h).
\end{equation}

Furthermore, the complement of the set \eqref{eq:2-001} can be defined as follows:

\begin{equation}
\Op_{x,\alpha}^{n,c}(h) := \left\{ o_x^\alpha \ : \ \exists o_k^\alpha h(x) \ \forall k \geq 1 \ \text{and} \ \lim_{\alpha \to n} o_k^\alpha h(x) \neq \partial_k^n h(x) \ \text{for at least one value of } k \right\}.
\end{equation}

From this, we obtain the following result:

\begin{equation}
\text{If } o_{i,x}^\alpha = \hat{e}_k o_{i,k}^\alpha \in \Op_{x,\alpha}^n(h) \ \Rightarrow \ \exists o_{j,x}^\alpha = \hat{e}_k o_{i,\sigma_j(k)}^\alpha \in \Op_{x,\alpha}^{n,c}(h),
\end{equation}
where $\sigma_j: \{1,2,\cdots,m\} \to \{1,2,\cdots,m\}$ is a permutation other than the identity. 

Next, the set \eqref{eq:2-001} generalizes elements of conventional calculus. For example, if $\gamma \in \mathbb{N}_0^m$ and $x \in \mathbb{R}^m$, we define the following multi-index notation:

\begin{equation}
\left\{
\begin{array}{c}
\begin{array}{ccc}
 \gamma!:= \ds\prod_{k=1}^m [\gamma]_k !,& \abs{\gamma}:= \ds \sum_{k=1}^m [\gamma]_k\vspace{0.1cm}, &  x^\gamma:= \ds \prod_{k=1}^m [x]_k^{[\gamma]_k}
\end{array} \\
\der{\partial}{x}{\gamma}:=       
\der{\partial}{[x]_1}{[\gamma]_1} 
\der{\partial}{[x]_2}{[\gamma]_2}\cdots \der{\partial}{[x]_m}{[\gamma]_m}  
\end{array}\right. .
\end{equation}

Consider a function $h: \Omega \subset \mathbb{R}^m \to \mathbb{R}$ and the fractional operator:

\begin{equation}
s_x^{\alpha\gamma}(o_x^\alpha) := o_1^{\alpha[\gamma]_1} o_2^{\alpha[\gamma]_2} \cdots o_m^{\alpha[\gamma]_m}.
\end{equation}

The following set of fractional operators is defined:

\begin{equation}
\Ops_{x,\alpha}^{n,\gamma}(h) := \left\{ s_x^{\alpha \gamma} = s_x^{\alpha \gamma}(o_x^\alpha) \ : \ \exists s_x^{\alpha \gamma} h(x) \ \text{with} \ o_x^\alpha \in \Op_{x,\alpha}^s(h), \ \forall s \leq n^2, \ \lim_{\alpha \to k} s_x^{\alpha \gamma} h(x) = \der{\partial}{x}{k\gamma} h(x), \ \forall \alpha, \abs{\gamma} \leq n \right\}.
\end{equation}

From this, the following results hold:

\begin{equation}
\mbox{If }s_x^{\alpha \gamma}\in \Ops_{x,\alpha}^{n,\gamma}(h) \ \Rightarrow \ \left\{
\begin{array}{l}
\ds \lim_{\alpha \to 0}s_x^{\alpha \gamma}h(x)=o_1^{0}o_2^{0}\cdots o_m^{0} h(x)=h(x) \vspace{0.1cm}\\
\ds \lim_{\alpha \to 1} s_x^{\alpha \gamma}h(x)= o_1^{[\gamma]_1}o_2^{[\gamma]_2}\cdots o_m^{[\gamma]_m} h(x)=\der{\partial}{x}{\gamma}h(x) \hspace{0.15cm} \forall \abs{\gamma}\leq n \vspace{0.1cm}\\
\ds \lim_{\alpha \to q} s_x^{\alpha \gamma}h(x)= o_1^{q[\gamma]_1}o_2^{q[\gamma]_2}\cdots o_m^{q[\gamma]_m} h(x)=\der{\partial}{x}{q\gamma}h(x) \hspace{0.15cm} \forall q\abs{\gamma}\leq qn 
\vspace{0.1cm}\\
\ds \lim_{\alpha \to n} s_x^{\alpha \gamma}h(x)= o_1^{n[\gamma]_1}o_2^{n[\gamma]_2}\cdots o_m^{n[\gamma]_m} h(x)=\der{\partial}{x}{n\gamma}h(x) \hspace{0.15cm} \forall n\abs{\gamma}\leq n^2 
\end{array}\right. .
\end{equation}

Using little-o notation, the following result can be derived:

\begin{equation}
\text{If } x \in B(a; \delta), \ \Rightarrow \ \lim_{x \to a} \frac{o((x-a)^\gamma)}{(x-a)^\gamma} \to 0, \quad \forall \abs{\gamma} \geq 1.
\end{equation}

This allows us to define the following set of functions:

\begin{equation}
R_{\alpha\gamma}^n(a) := \left\{ r_{\alpha\gamma}^n \ : \ \lim_{x \to a} \| r_{\alpha\gamma}^n(x) \| = 0, \ \forall \abs{\gamma} \geq n, \ \| r_{\alpha\gamma}^n(x) \| \leq o(\| x - a \|^n), \ \forall x \in B(a; \delta) \right\}.
\end{equation}

Now, consider the following sets of fractional operators:

\begin{equation}
\Opt_{x,\alpha,p}^{n,q,\gamma}(a,h) := \left\{ t_x^{\alpha,p} = t_x^{\alpha,p}(s_x^{\alpha \gamma}) \ : \ s_x^{\alpha \gamma} \in \Ops_{x,\alpha}^{M,\gamma}(h), \ t_x^{\alpha,p} h(x) := \sum_{\abs{\gamma}=0}^p \frac{1}{\gamma!} s_x^{\alpha \gamma} h(a) (x-a)^\gamma + r_{\alpha\gamma}^p(x) \ \forall \alpha \leq n, \forall p \leq q \right\},
\end{equation}

\begin{equation}
\Opt_{x,\alpha}^{\infty,\gamma}(a,h) := \left\{ t_x^{\alpha,\infty} = t_x^{\alpha,\infty}(s_x^{\alpha \gamma}) \ : \ s_x^{\alpha \gamma} \in \Ops_{x,\alpha}^{\infty,\gamma}(h), \ t_x^{\alpha,\infty} h(x) := \sum_{\abs{\gamma}=0}^\infty \frac{1}{\gamma!} s_x^{\alpha \gamma} h(a) (x-a)^\gamma \right\}.
\end{equation}

These sets generalize the Taylor series expansion of a scalar function in multi-index notation~\cite{torres2021fracnewrapaitken}, where $M = \max\{n, q\}$.

Consequently, the following results hold:

\begin{equation}
\text{If } t_x^{\alpha,p} \in \Opt_{x,\alpha,p}^{1,q,\gamma}(a,h) \text{ and } \alpha \to 1, \ \Rightarrow \ t_x^{1,p} h(x) = h(a) + \sum_{\abs{\gamma}=1}^p \frac{1}{\gamma!} \der{\partial}{x}{\gamma} h(a) (x-a)^\gamma + r_{\gamma}^p(x).
\end{equation}

\begin{equation}
\text{If } t_x^{\alpha,p} \in \Opt_{x,\alpha,p}^{n,1,\gamma}(a,h) \text{ and } p \to 1, \ \Rightarrow \ t_x^{\alpha,1} h(x) = h(a) + \sum_{k=1}^m o_k^\alpha h(a) \left[(x-a)\right]_k + r_{\alpha\gamma}^1(x).
\end{equation}

Finally, the set \eqref{eq:2-001} can be seen as a generating set for fractional tensor operators. For example, if $\alpha, n \in \mathbb{R}^d$ with $\alpha = \hat{e}_k[\alpha]_k$ and $n = \hat{e}_k [n]_k$, the following set of fractional tensor operators can be defined:

\begin{equation}
\Op_{x,\alpha}^{n}(h) := \left\{ o_x^\alpha \ : \ \exists o_x^\alpha h(x) \ \text{and} \ o_x^\alpha \in \Op_{x,[\alpha]_1}^{[n]_1}(h) \times \Op_{x,[\alpha]_2}^{[n]_2}(h) \times \cdots \times \Op_{x,[\alpha]_d}^{[n]_d}(h) \right\}.
\end{equation}

\section{Groups of Fractional Operators}

Let \( h: \Omega \subset \mathbb{R}^m \to \mathbb{R}^m \) be a function. We define the sets of fractional operators for a vector function as follows:

\begin{gather}
\mspc{{}_m\Op_{x,\alpha}^n(h):=\set{ o_x^\alpha \ : \ o_x^\alpha\in \Op_{x,\alpha}^n\left([h]_k \right) \ \forall k \leq m},}{0.1cm}\\
\mspc{
{}_m\Op_{x,\alpha}^{n,c}(h):=\set{ o_x^\alpha \ : \ o_x^\alpha\in \Op_{x,\alpha}^{n,c}\left([h]_k \right) \ \forall k \leq m},}{0.1cm}\\
{}_m\Op_{x,\alpha}^{n,u}(h):={}_m\Op_{x,\alpha}^{n}(h)\cup{}_m\Op_{x,\alpha}^{n,c}(h),
\end{gather}

where \( [h]_k: \Omega \subset \mathbb{R}^m \to \mathbb{R} \) is the \( k \)-th component of \( h \). Using these sets, we can define the following family of fractional operators:

\begin{equation}
{}_m\MOp_{x,\alpha}^{\infty,u}(h):=\bigcap_{k\in \nset{Z}}{}_m\Op_{x,\alpha}^{k,u}(h).
\end{equation}

It is important to note that this family of fractional operators satisfies the following property with respect to the classical Hadamard product:
\begin{equation}
o_x^0 \circ h(x):= h(x) \quad \forall o_x^\alpha \in {}_m\MOp_{x,\alpha}^{\infty,u}(h).
\end{equation}

For each operator \( o_x^\alpha \in \mathcal{M}\mathcal{O}_{x,\alpha}^{\infty,u}(h) \), we can define the fractional matrix operator \cite{torres2022codeaccelfracquasi}:
\begin{equation}
A_\alpha(o_x^\alpha) = \left( [A_\alpha(o_x^\alpha)]_{jk} \right) := (o_k^\alpha).
\end{equation}

Next, we define a modified Hadamard product \cite{torres2021sets}:
\begin{equation}\label{eq:3-001}
o_{i,x}^{p\alpha} \circ o_{j,x}^{q\alpha}:=
\begin{cases}
o_{i,x}^{p\alpha} \circ o_{j,x}^{q\alpha}, & \text{if } i \neq j \text{ (horizontal Hadamard product)} \\
o_{i,x}^{(p+q)\alpha}, & \text{if } i = j \text{ (vertical Hadamard product)}
\end{cases}
\end{equation}
for each operator \( o_x^\alpha \in \MOp_{x,\alpha}^{\infty,u}(h) \). This enables us to define an Abelian group of fractional operators isomorphic to the group of integers under addition, as shown by the following theorem \cite{torres2022codeaccelfracquasi,torres2022acceleration}:

\begin{theorem}
Let \( o_x^\alpha \) be a fractional operator such that \( o_x^\alpha \in \MOp_{x,\alpha}^{\infty,u}(h) \), and let \( (\mathbb{Z}, +) \) be the group of integers under addition. Then, using the modified Hadamard product defined by \eqref{eq:3-001}, we define the set of fractional matrix operators:
\begin{equation}
\Og_m(A_\alpha(o_x^\alpha)) := \{ A_\alpha^{\circ r} = A_\alpha(o_x^{r\alpha}) : r \in \mathbb{Z} \},
\end{equation}
which corresponds to the Abelian group generated by the operator \( A_\alpha(o_x^\alpha) \), isomorphic to the group \( (\mathbb{Z}, +) \), i.e.,
\begin{equation}
\Og_m(A_\alpha(o_x^\alpha)) \cong (\mathbb{Z}, +).
\end{equation}

\begin{proof}
The set \( \Og_m(A_\alpha(o_x^\alpha)) \) is closed under the modified vertical Hadamard product. For all \( A_\alpha^{\circ p}, A_\alpha^{\circ q} \in \Og_m(A_\alpha(o_x^\alpha)) \), we have:
\begin{equation}
A_\alpha^{\circ p} \circ A_\alpha^{\circ q} = A_\alpha^{\circ(p+q)}.
\end{equation}

Thus, the set forms a semigroup:
\begin{equation}
\forall A_\alpha^{\circ p}, A_\alpha^{\circ q}, A_\alpha^{\circ r} \in \Og_m(A_\alpha(o_x^\alpha)), \quad (A_\alpha^{\circ p} \circ A_\alpha^{\circ q}) \circ A_\alpha^{\circ r} = A_\alpha^{\circ p} \circ (A_\alpha^{\circ q} \circ A_\alpha^{\circ r}).
\end{equation}

The set also contains an identity element, making it a monoid:
\begin{equation}
\exists A_\alpha^{\circ 0} \in \Og_m(A_\alpha(o_x^\alpha)), \quad A_\alpha^{\circ 0} \circ A_\alpha^{\circ p} = A_\alpha^{\circ p}.
\end{equation}

The symmetric element for each element in the set exists, making it a group:
\begin{equation}
\exists A_\alpha^{\circ -p} \in \Og_m(A_\alpha(o_x^\alpha)), \quad A_\alpha^{\circ p} \circ A_\alpha^{\circ -p} = A_\alpha^{\circ 0}.
\end{equation}

Finally, since the order of operation does not affect the result, the set is Abelian:
\begin{equation}
A_\alpha^{\circ p} \circ A_\alpha^{\circ q} = A_\alpha^{\circ q} \circ A_\alpha^{\circ p}.
\end{equation}

Thus, the set \( \Og_m(A_\alpha(o_x^\alpha)) \) is an Abelian group. To complete the proof, we define a bijective homomorphism \( \psi \) between \( \Og_m(A_\alpha(o_x^\alpha)) \) and \( (\mathbb{Z}, +) \):
\begin{equation}
\psi(A_\alpha^{\circ r}) = r, \quad \psi^{-1}(r) = A_\alpha^{\circ r}.
\end{equation}

\end{proof}
\end{theorem}

From the previous theorem, we obtain the following corollary:

\begin{corollary}\label{cor:3-001}
Let \( o_x^\alpha \) be a fractional operator such that \( o_x^\alpha \in \MOp_{x,\alpha}^{\infty,u}(h) \). Let \( (\mathbb{Z}, +) \) be the group of integers under addition, and \( \mathbb{H} \) a subgroup of \( \mathbb{Z} \). The set of fractional matrix operators is given by:
\begin{equation}
\Og_m(A_\alpha(o_x^\alpha), \mathbb{H}) := \{ A_\alpha^{\circ r} = A_\alpha(o_x^{r\alpha}) : r \in \mathbb{H} \}.
\end{equation}
This forms a subgroup of the group generated by \( A_\alpha(o_x^\alpha) \):
\begin{equation}
\Og_m(A_\alpha(o_x^\alpha), \mathbb{H}) \leq \Og_m(A_\alpha(o_x^\alpha)).
\end{equation}
\end{corollary}

\begin{example}
Let \( \mathbb{Z}_n \) be the set of residues modulo a positive integer \( n \). Given a fractional operator \( o_x^\alpha \in \MOp_{x,\alpha}^{\infty,u}(h) \) and \( \mathbb{Z}_{14} \), the following Abelian group of fractional matrix operators is defined under the modified Hadamard product:

\begin{equation}
\Og_m(A_\alpha(o_x^\alpha), \mathbb{Z}_{14}) = \set{A_\alpha^{\circ 0}     , A_\alpha^{\circ 1}     , A_\alpha^{\circ 2}     , A_\alpha^{\circ 3}     , A_\alpha^{\circ 4}     , A_\alpha^{\circ 5}     , A_\alpha^{\circ 6}     , A_\alpha^{\circ 7}     , A_\alpha^{\circ 8}     , A_\alpha^{\circ 9}     , A_\alpha^{\circ 10}    , A_\alpha^{\circ 11}    , A_\alpha^{\circ 12}    , A_\alpha^{\circ 13}}.
\end{equation}
\end{example}


It is important to note that Corollary~\ref{cor:3-001} allows generating groups of fractional operators under different operations. For instance, given the operation
\begin{eqnarray}\label{eq:3-003}
A_{\alpha}^{\circ r}*A_{\alpha}^{\circ s}=A_{\alpha}^{\circ rs},
\end{eqnarray}
we can derive the following corollaries:

\begin{corollary}
Let $\nset{M}_n$ represent the set of positive residual classes corresponding to the coprimes less than a positive integer $n$. For each fractional operator $o_x^\alpha \in {}_m\MOp_{x,\alpha}^{\infty,u}(h)$, we can define an Abelian group of fractional matrix operators under the operation \eqref{eq:3-003}:
\begin{eqnarray}
{}_m\Og^*\left(A_\alpha\left(o_x^\alpha \right), \nset{M}_n \right) := \left\{ A_\alpha^{\circ r} = A_\alpha\left(o_x^{r\alpha}\right) : r \in \nset{M}_n \right\}.
\end{eqnarray}
\end{corollary}

\begin{example}
Let $o_x^\alpha$ be a fractional operator such that $o_x^\alpha \in {}_m\MOp_{x,\alpha}^{\infty,u}(h)$. For the set $\nset{M}_{14}$, we can define the Abelian group of fractional matrix operators as follows under the operation \eqref{eq:3-003}:
\begin{eqnarray}
{}_m\Og^*\left(A_\alpha\left(o_x^\alpha \right), \nset{M}_{14} \right) = \{ A_\alpha^{\circ 1}, A_\alpha^{\circ 3}, A_\alpha^{\circ 5}, A_\alpha^{\circ 9}, A_\alpha^{\circ 11}, A_\alpha^{\circ 13} \}.
\end{eqnarray}
\end{example}

\begin{corollary}
Let $\nset{Z}_p^+$ represent the set of positive residual classes less than a prime $p$. For each fractional operator $o_x^\alpha \in {}_m\MOp_{x,\alpha}^{\infty,u}(h)$, we can define the Abelian group of fractional matrix operators under the operation \eqref{eq:3-003} as:
\begin{eqnarray}
{}_m\Og^*\left(A_\alpha\left(o_x^\alpha \right), \nset{Z}_p^+ \right) := \left\{ A_\alpha^{\circ r} = A_\alpha\left(o_x^{r\alpha}\right) : r \in \nset{Z}_p^+ \right\}.
\end{eqnarray}
\end{corollary}

\begin{example}
Let $o_x^\alpha$ be a fractional operator such that $o_x^\alpha \in {}_m\MOp_{x,\alpha}^{\infty,u}(h)$. For the set $\nset{Z}_{13}^+$, we can define the Abelian group of fractional matrix operators as:
\begin{eqnarray}
{}_m\Og^*\left(A_\alpha\left(o_x^\alpha \right), \nset{Z}_{13}^+ \right) = \{ A_\alpha^{\circ 1}, A_\alpha^{\circ 2}, A_\alpha^{\circ 3}, A_\alpha^{\circ 4}, A_\alpha^{\circ 5}, A_\alpha^{\circ 6}, A_\alpha^{\circ 7}, A_\alpha^{\circ 8}, A_\alpha^{\circ 9}, A_\alpha^{\circ 10}, A_\alpha^{\circ 11}, A_\alpha^{\circ 12} \}.
\end{eqnarray}
\end{example}

Finally, when $n$ is a prime number, the following result holds:
\begin{eqnarray}
{}_m\Og^*\left(A_\alpha\left(o_x^\alpha \right), \nset{M}_n \right) = {}_m\Og^*\left(A_\alpha\left(o_x^\alpha \right), \nset{Z}_n^+ \right).
\end{eqnarray}

\section{Polynomial Regression Model}

Polynomial regression is an extension of linear regression that models the relationship between the independent variable $x$ and the dependent variable $y$ using a polynomial function of degree $n$. Unlike simple linear regression, which assumes a strictly linear relationship between $x$ and $y$, polynomial regression allows for curvature by incorporating higher-order terms of $x$. The general form of a polynomial regression model is given by:

\begin{eqnarray*}
y = \beta_0 + \beta_1 x + \beta_2 x^2 + \dots + \beta_n x^n.
\end{eqnarray*}

In this expression, the term $\beta_0$ represents the intercept, which is the expected value of $y$ when $x = 0$. The coefficients $\beta_1, \beta_2, \dots, \beta_n$ determine the contribution of each corresponding power of $x$ to the predicted value of $y$. 

The inclusion of higher-degree terms allows polynomial regression to model nonlinear trends while still maintaining a linear structure with respect to the parameters. The flexibility of the model increases as the degree of the polynomial $n$ grows, enabling it to approximate complex relationships in data.

The polynomial regression model can also be expressed using matrix notation, which facilitates the estimation of the coefficients. In matrix form, the model is written as:

\begin{eqnarray*}
\mathbf{y} = \mathbf{X} \boldsymbol{\beta},
\end{eqnarray*}

where:

\begin{itemize}
    \item $\mathbf{y} = \begin{bmatrix} y_1 \\ y_2 \\ \vdots \\ y_m \end{bmatrix}$ is the vector of observed values,
    \item $\mathbf{X} = \begin{bmatrix} 
        1 & x_1 & x_1^2 & \dots & x_1^n \\ 
        1 & x_2 & x_2^2 & \dots & x_2^n \\ 
        \vdots & \vdots & \vdots & \ddots & \vdots \\ 
        1 & x_m & x_m^2 & \dots & x_m^n 
    \end{bmatrix}$ is the design matrix, where each row corresponds to an observation and each column represents a different power of $x$,
    \item $\boldsymbol{\beta} = \begin{bmatrix} \beta_0 \\ \beta_1 \\ \vdots \\ \beta_n \end{bmatrix}$ is the vector of coefficients.
\end{itemize}

The estimation of $\boldsymbol{\beta}$ is typically performed using the least squares method, which provides the best-fitting polynomial by minimizing the differences between observed and predicted values. The solution for $\boldsymbol{\beta}$ is given by:

\begin{eqnarray*}
\hat{\boldsymbol{\beta}} = (\mathbf{X}^T \mathbf{X})^{-1} \mathbf{X}^T \mathbf{y}.
\end{eqnarray*}

This equation represents a fundamental aspect of polynomial regression, as it allows the determination of the coefficients that best describe the observed data.

Each coefficient in the polynomial regression model has a specific role in shaping the regression curve:

\begin{itemize}
    \item The intercept $\beta_0$ represents the predicted value of $y$ when $x = 0$.
    \item The coefficient $\beta_1$ is associated with the linear term $x$, determining the slope of the regression line at $x=0$.
    \item The coefficient $\beta_2$ is associated with the quadratic term $x^2$ and influences the curvature of the function.
    \item Higher-order coefficients, such as $\beta_3$ for $x^3$, $\beta_4$ for $x^4$, and so on, add additional flexibility, allowing the model to capture more complex patterns.
\end{itemize}

The choice of polynomial degree $n$ is a critical factor in modeling. A low-degree polynomial may fail to capture important trends in the data, whereas a high-degree polynomial can lead to excessive fluctuations, resulting in an overly complex model. The selection of an appropriate degree often involves techniques such as cross-validation and domain knowledge.

Polynomial regression has numerous applications across various fields:

\begin{itemize}
    \item \textbf{Economics:} Used to model trends in financial markets, cost functions, and consumer behavior.
    \item \textbf{Physics:} Applied in trajectory modeling, wave analysis, and thermodynamics to describe nonlinear phenomena.
    \item \textbf{Engineering:} Utilized in control systems, signal processing, and material science for curve fitting and performance modeling.
    \item \textbf{Machine Learning:} Employed in feature transformation and basis function expansion to enhance the representation of complex relationships in data.
\end{itemize}

By allowing for nonlinear relationships, polynomial regression extends the power of classical regression models while maintaining a simple and interpretable structure based on coefficients and intercepts.

\section{Seen and Unseen Data in Regression Models}

In regression analysis, understanding the distinction between "seen" and "unseen" data is fundamental to evaluating the performance and generalization ability of a model. This distinction directly influences how well a model can adapt to new, unseen data and whether it can produce reliable predictions when applied to real-world scenarios.

\textbf{Seen data}, also referred to as \textit{training data}, represents the dataset used by the model during its training phase. During this phase, the model learns the underlying patterns and relationships between the independent variables (features) and the dependent variable (target). By adjusting its parameters, such as coefficients, the model seeks to minimize the error between its predictions and the actual values within the training set. The quality and quantity of the training data play a significant role in determining how accurately the model can represent the relationships between features and target variables.

The main objective of training a model on seen data is to create a function that describes the relationship between the input features and the output target. However, the model must not only fit the training data well, but also generalize its learned patterns to future, unseen data. This generalization ability is critical, as it determines how useful the model will be when faced with new datasets or in practical applications.

\textbf{Unseen data}, also known as \textit{test data}, refers to data that the model has never encountered during the training process. This dataset is used to assess how well the model performs when exposed to new information that it hasn’t been trained on. The primary goal of using unseen data is to evaluate the model's predictive accuracy and to check whether it has successfully learned the underlying patterns without memorizing the training data.

The ability of a model to generalize to unseen data is often referred to as \textit{generalization}. A model that generalizes well can apply its learned knowledge to make accurate predictions about new data. This is a hallmark of a robust model. If the model performs well on both seen and unseen data, it suggests that it has captured the true underlying relationships in the data rather than simply fitting to noise or anomalies in the training set. On the other hand, if the model performs poorly on unseen data, it may be an indication that it has failed to generalize properly, potentially due to overfitting or underfitting.

\begin{itemize}
    \item \textbf{Overfitting:} Occurs when the model learns the intricacies and noise of the training data to such an extent that it negatively impacts its performance on unseen data. Overfitting is often a result of a model that is too complex relative to the amount or variability of the training data. This means that the model becomes overly specialized to the training data and loses its ability to generalize to new, unseen examples.
    
    \item \textbf{Underfitting:} Occurs when the model is too simplistic to capture the true patterns within the data, leading to poor performance on both the training and test sets. Underfitting typically happens when the model has too few parameters or is too rigid to learn the underlying relationships in the data, resulting in both inaccurate predictions and low model performance.
\end{itemize}

\section{The Coefficient of Determination \( R^2 \)}

One of the key metrics used to evaluate the performance of a regression model is the coefficient of determination, denoted as \( R^2 \). This metric measures how well the independent variables explain the variance of the dependent variable. Mathematically, \( R^2 \) is defined as:

\begin{eqnarray*}
R^2 = 1 - \frac{\sum (y_i - \hat{y}_i)^2}{\sum (y_i - \bar{y})^2},
\end{eqnarray*}

where \( y_i \) represents the actual values, \( \hat{y}_i \) are the predicted values, and \( \bar{y} \) is the mean of the observed values. The numerator quantifies the residual sum of squares (the error between predictions and actual values), while the denominator represents the total sum of squares (the variance in the data). A higher \( R^2 \) value, closer to 1, indicates that a greater proportion of variance in the dependent variable is explained by the model.

The coefficient of determination \( R^2 \) can take values between 0 and 1, although negative values are possible in certain situations, particularly when the model is poorly specified or overfitting. The following outlines the interpretation of different \( R^2 \) values:

\begin{itemize}
    \item \textbf{\( R^2 = 1: \)} A value of \( R^2 = 1 \) means that the model perfectly explains the variance in the dependent variable. In other words, the predicted values match the actual values exactly, and there is no error in the model’s predictions. However, achieving this in real-world scenarios is rare, and a value of 1 often indicates potential overfitting.
    
    \item \textbf{\( 0 < R^2 < 1: \)} Values of \( R^2 \) between 0 and 1 indicate that the model explains a portion of the variance in the dependent variable. The closer \( R^2 \) is to 1, the better the model fits the data and the more variance it explains. For example, an \( R^2 \) of 0.8 suggests that 80
    
    \item \textbf{\( R^2 = 0: \)} An \( R^2 \) of 0 means that the model does not explain any of the variance in the dependent variable. In other words, the model's predictions are no better than simply predicting the mean value of the dependent variable for all observations. This indicates that the model is not capturing any useful relationships between the features and the target variable.
    
    \item \textbf{Negative \( R^2 \):} While unusual, it is possible to obtain a negative \( R^2 \), which typically happens when the model is worse than a simple model that just predicts the mean value of the dependent variable. A negative \( R^2 \) suggests that the model has a poor fit and is performing worse than random guessing. This often occurs in cases of overfitting or model mis-specification.
\end{itemize}

In general, a higher \( R^2 \) indicates better model performance and the ability to explain the variance in the target variable. However, \( R^2 \) alone should not be used as the sole metric for evaluating model performance, as it does not account for overfitting or the complexity of the model. It is important to consider additional metrics and validation techniques, such as cross-validation, to ensure the robustness of the model.

While a high \( R^2 \) on training data suggests a good fit, it is even more crucial to achieve a strong \( R^2 \) on unseen test data. A high \( R^2 \) on unseen data signifies that the model is capturing the true relationships within the dataset, rather than overfitting to the training data. This has several important advantages:

\begin{itemize}
    \item \textbf{Better Generalization:} A model with a high \( R^2 \) on unseen data is more reliable for making predictions on new data, making it more useful in real-world applications.
    \item \textbf{Reduced Overfitting Risk:} If the model maintains a strong \( R^2 \) on test data, it suggests that it has not simply memorized training data but has genuinely learned meaningful patterns.
    \item \textbf{Improved Decision-Making:} When applied in fields like finance, medicine, and engineering, a model with a strong \( R^2 \) on unseen data provides more trustworthy predictions, aiding in better decision-making.
    \item \textbf{More Robust Model Selection:} Comparing \( R^2 \) across different models helps in selecting the best-performing model that balances complexity and predictive power without overfitting.
\end{itemize}

Thus, ensuring that the model maintains a high \( R^2 \) not just on training data but also on unseen test data is essential for developing reliable and interpretable regression models.

\section{Fractional Regression Model}

When generating a predictive model based on time series, the initial strategy typically follows a common approach: dividing the data into two main parts. The first part, corresponding to interpolation, includes the data within the training set. The second part, related to extrapolation, includes the data located at the extremes of the set, and its function is to validate the model's predictions once the training and testing phases are completed.

In polynomial regression models, when setting a value $m$ for the polynomial order, if the result in the validation set is not as expected, variations can be excessive when modifying this order by one unit, either up or down. These changes, which can sometimes be very radical, lead to a rethinking of the approaches adopted in some projects. Due to this phenomenon, this work proposes the partial implementation of fractional derivatives in regression models, based on the works published in \cite{torres2023proposal, torres2021blackscholes}. To test this proposal, polynomial regression models are generated using a dataset containing the average prices of conventional and organic avocados between 2015 and 2018, available in the following GitHub repository: \href{https://github.com/UOC-curso/curso/blob/main/proyectos/proyecto-1-regresion/avocado.csv}{avocado.csv}.

Denoting a polynomial regression model of order $m$ as follows:

\begin{eqnarray}
    y(x) = \sum_{i=0}^m \beta_i x^i.
\end{eqnarray}

where it is assumed that the values of $\{\beta_i\}_{i=0}^m$ have been found for some value of $m$. Then, consider the following set:

\begin{eqnarray}
    \MOp_{x,\alpha}^{m}(y):=\bigcap_{k=-m}^{m}\Op_{x,\alpha}^{k}(y),
\end{eqnarray}

it is possible to define the following fractional regression model for any operator $o_x^\alpha \in \MOp_{x,\alpha}^{1}(y)$:

\begin{eqnarray}
    \sigma(\alpha, x):= \beta_0 + \sum_{i=1}^m \beta_i o_x^\alpha x^i \quad \text{with} \quad \alpha \in (-1,1).
\end{eqnarray}

It is important to note that the intercept value must be kept intact to respect the value of the polynomial regression model in the case of $x=0$. On the other hand, if a multidimensional space and a logistic function denoted by $f$ are considered, the previous equation would also allow defining a fractional logistic regression model as follows:  

\begin{eqnarray}
    \sigma(\alpha, x):= f\left( \beta_0 + \sum_{i=1}^m \beta_i o_{x_i}^\alpha x_i\right)  \quad \text{with} \quad \alpha \in (-1,1).
\end{eqnarray}

This, in principle, would generate a fractional activation function, opening the possibility to develop a fractional neural network, provided that the characteristics $x_i$ are non-negative. To ensure this condition, a normalization process could be applied to the features, ensuring that the minimum value of each one is 0 and the maximum is 1.

\subsection{Examples Using Fractional Regression Model }

Considering that for any polynomial regression model $y$ of degree $m$, without considering the intercept and a value of $\alpha \in (-1,1)$, the Riemann-Liouville and Caputo fractional operators given by equations \eqref{eq:1-001} and \eqref{eq:1-002} belong to the set $\MOp_{x,\alpha}^{1}(y)$ and satisfy equation \eqref{eq:1-003}. Then, the following examples can be considered valid for both fractional operators using equation \eqref{eq:3-002}.

Taking the average price values of conventional avocado from the dataset, a region is selected, and the values are grouped monthly. Since some groupings show considerable dispersion, the value of the \textbf{monthly median} in each group is chosen to obtain a point distribution that is less sensitive to extreme values compared to the one obtained using the monthly mean.

For this point distribution, a split into two sets is made:

\begin{itemize}
    \item \textbf{Interpolation Part}: Contains $80\%$ of the data.
    \item \textbf{Extrapolation Part}: Contains the remaining $20\%$ of the data.
\end{itemize}

This split is achieved through the following function in \textit{Python}:

\begin{eqnarray*}
\texttt{train\_test\_split}(X, \texttt{y\_price}, \texttt{test\_size}=0.2, \texttt{random\_state}=42, \texttt{shuffle}=\texttt{False})
\end{eqnarray*}

The function \texttt{train\_test\_split} from the \texttt{sklearn.model\_selection} library allows splitting a dataset into two subsets: one for training (in this case, interpolation) and the other for testing (in this case, extrapolation). 

\begin{itemize}
    \item \textbf{X}: Input dataset, in this case, the time or date values.
    \item \textbf{y\_price}: Values corresponding to the average price of conventional avocado.
    \item \textbf{test\_size=0.2}: Indicates that $20\%$ of the data will be used for extrapolation.
    \item \textbf{random\_state=42}: Sets the seed to ensure reproducibility of the results.
    \item \textbf{shuffle=False}: Indicates that the data should not be shuffled before the split, preserving the temporal order.
\end{itemize}

This procedure allows generating a dataset for evaluating polynomial prediction models and the subsequent construction of fractional regression models in the analysis of average conventional avocado prices by region.

\begin{figure}[!ht]
    \centering
    \subfloat[Box plots of average prices grouped by month]{%
        \includegraphics[width=0.48\textwidth]{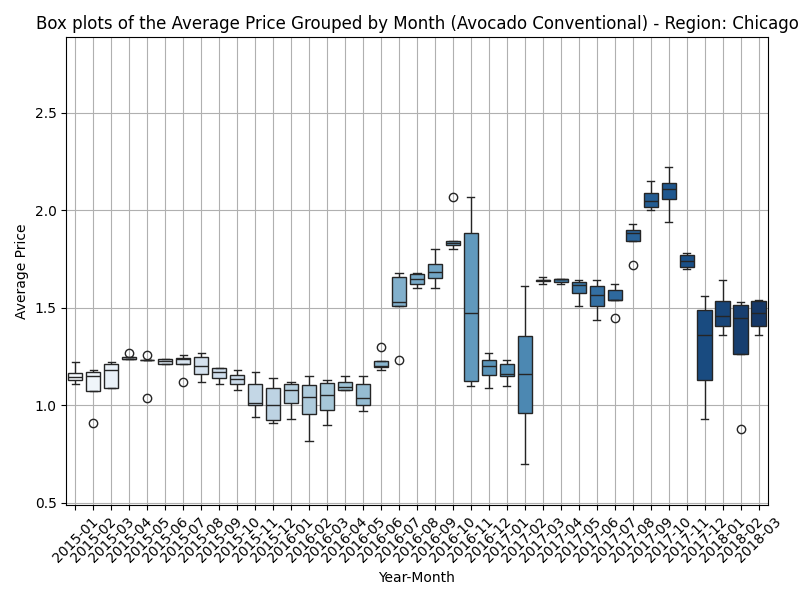}%
    }
    \hfill
    \subfloat[Logarithmic scale graph of positive $R^2$ values for interpolation and extrapolation parts simultaneously]{%
        \includegraphics[width=0.48\textwidth]{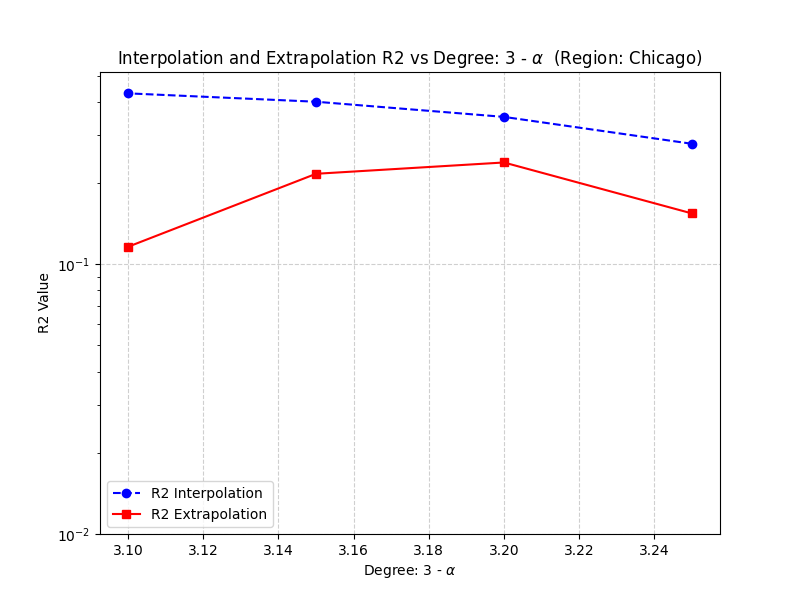}%
    }
    
    \vspace{0.1cm} 

    \subfloat[Average price distribution using the monthly median - Fractional regression model with $\alpha = 0$]{%
        \includegraphics[width=0.48\textwidth]{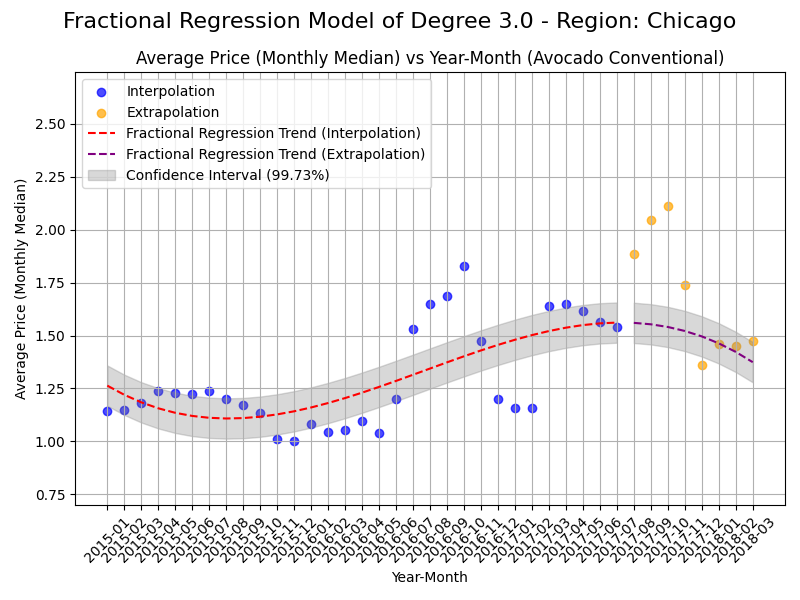}%
    }
    \hfill
    \subfloat[Average price distribution using the monthly median - Fractional regression model with $\alpha \neq 0$]{%
        \includegraphics[width=0.48\textwidth]{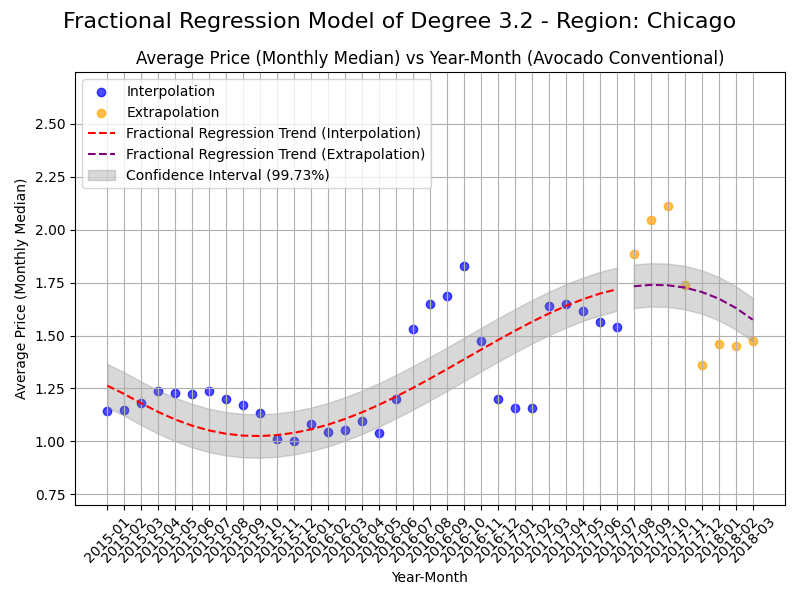}%
    }

    \caption{Comparison of metrics and fractional regression models in the Chicago region}
    \label{fig:chicago_analysis}
\end{figure}

\begin{figure}[!ht]
    \centering
    \subfloat[Box plots of average prices grouped by month]{%
        \includegraphics[width=0.48\textwidth]{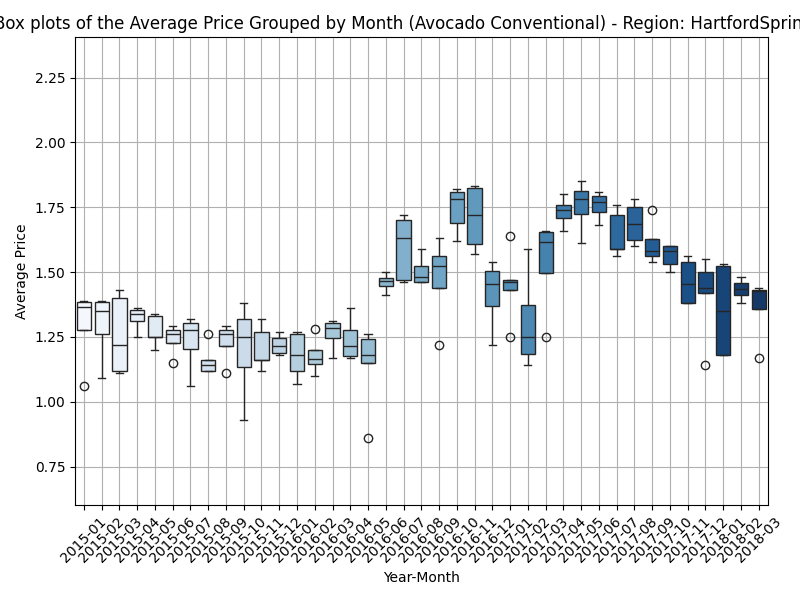}%
    }
    \hfill
    \subfloat[Logarithmic scale graph of positive $R^2$ values for interpolation and extrapolation parts simultaneously]{%
        \includegraphics[width=0.48\textwidth]{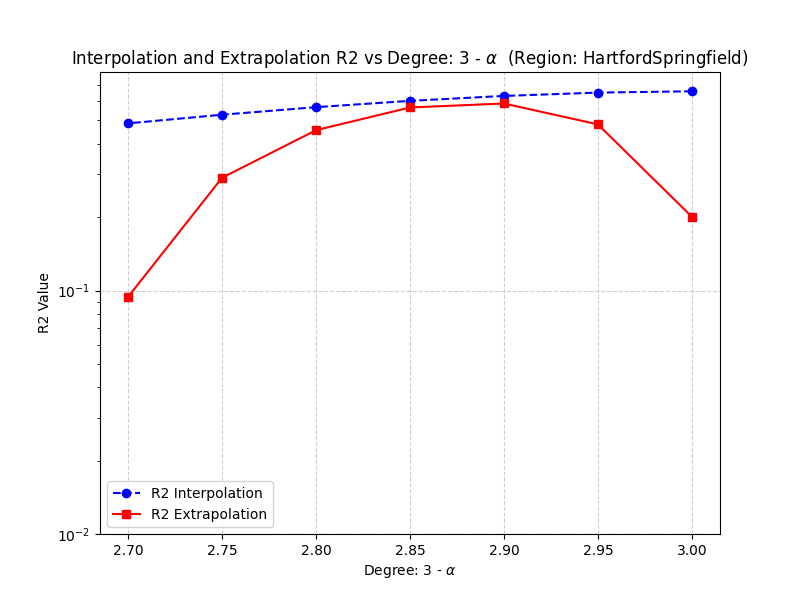}%
    }
    
    \vspace{0.1cm} 

    \subfloat[Average price distribution using the monthly median - Fractional regression model with $\alpha = 0$]{%
        \includegraphics[width=0.48\textwidth]{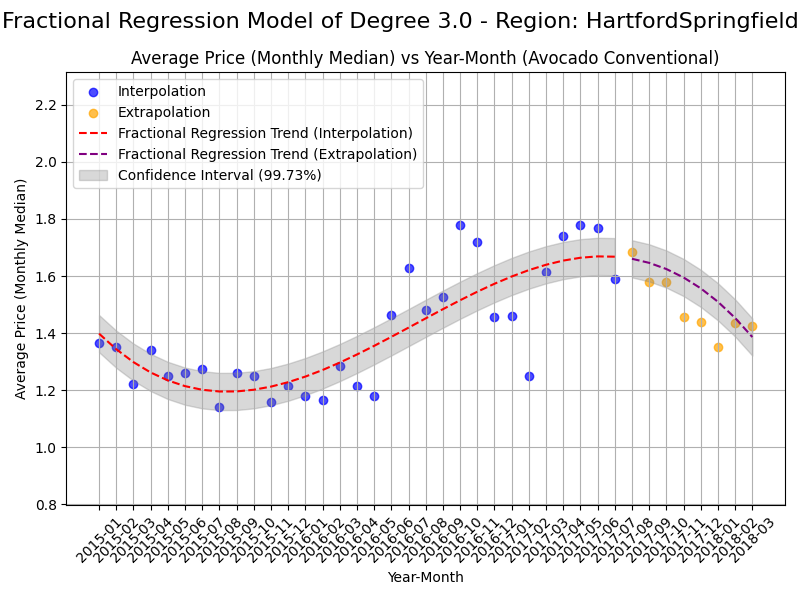}%
    }
    \hfill
    \subfloat[Average price distribution using the monthly median - Fractional regression model with $\alpha \neq 0$]{%
        \includegraphics[width=0.48\textwidth]{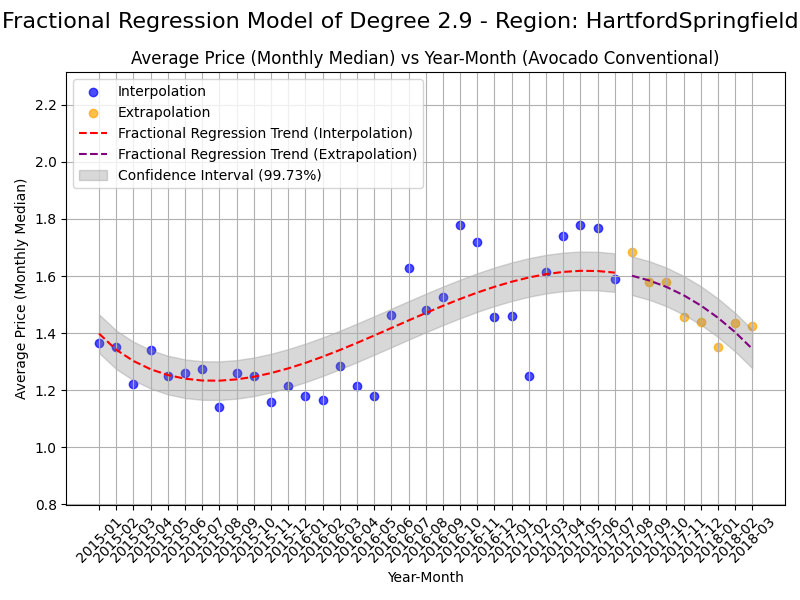}%
    }

    \caption{Comparison of metrics and fractional regression models in the HartfordSpringfield region }
    \label{fig:hartford_analysis}
\end{figure}

\begin{figure}[!ht]
    \centering
    \subfloat[Box plots of average prices grouped by month]{%
        \includegraphics[width=0.48\textwidth]{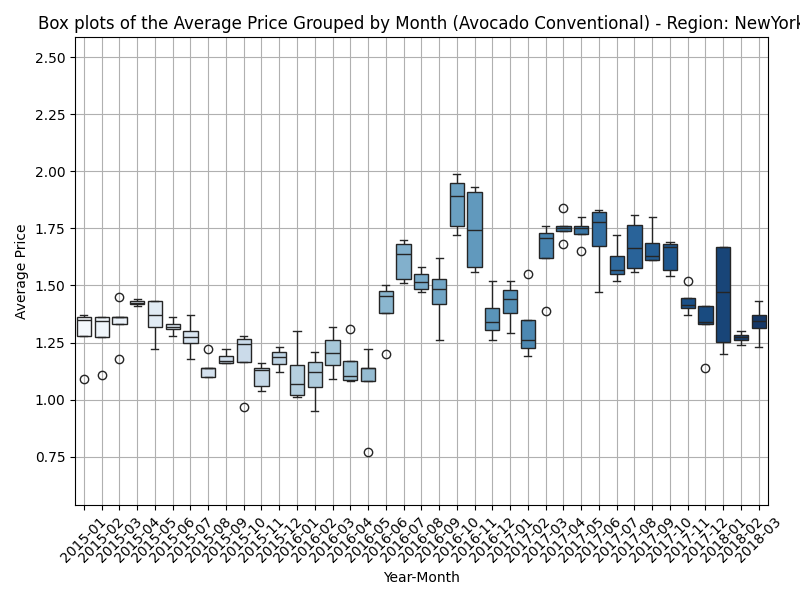}%
    }
    \hfill
    \subfloat[Logarithmic scale graph of positive $R^2$ values for interpolation and extrapolation parts simultaneously]{%
        \includegraphics[width=0.48\textwidth]{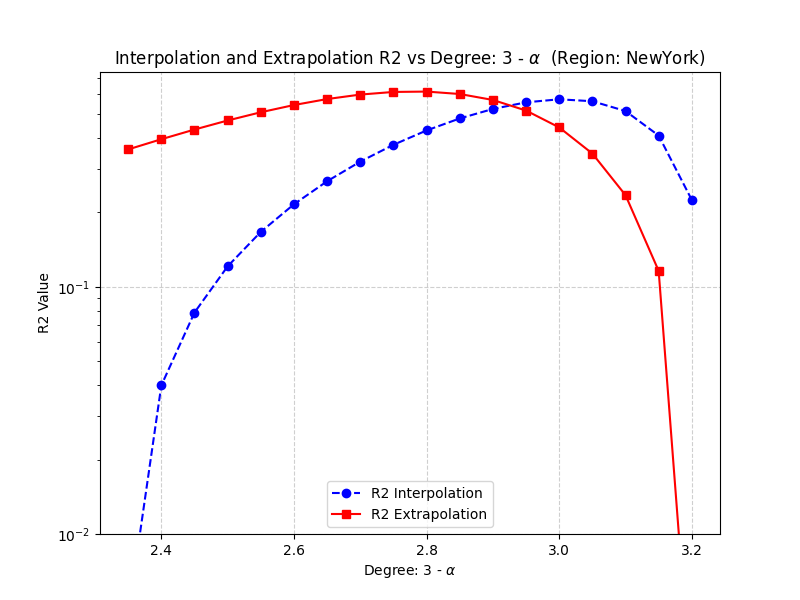}%
    }
    
    \vspace{0.1cm} 

    \subfloat[Average price distribution using the monthly median - Fractional regression model with $\alpha = 0$]{%
        \includegraphics[width=0.48\textwidth]{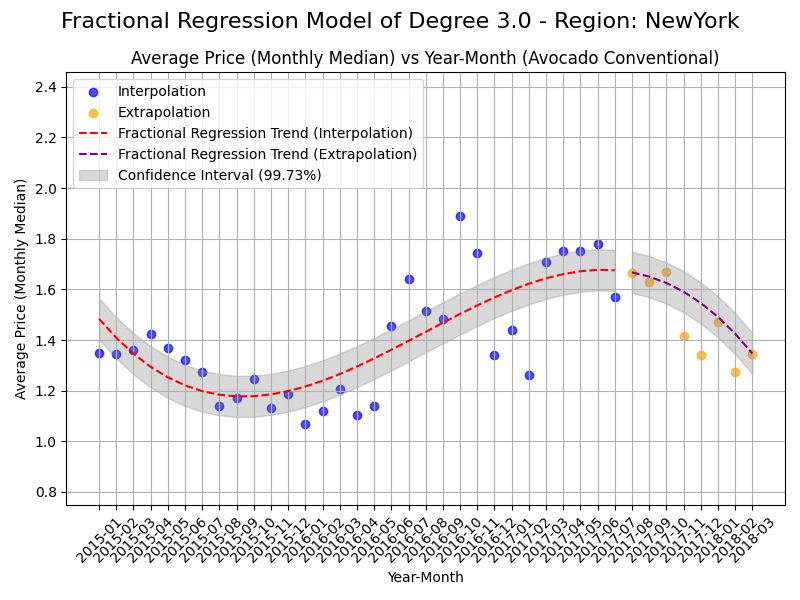}%
    }
    \hfill
    \subfloat[Average price distribution using the monthly median - Fractional regression model with $\alpha \neq 0$]{%
        \includegraphics[width=0.48\textwidth]{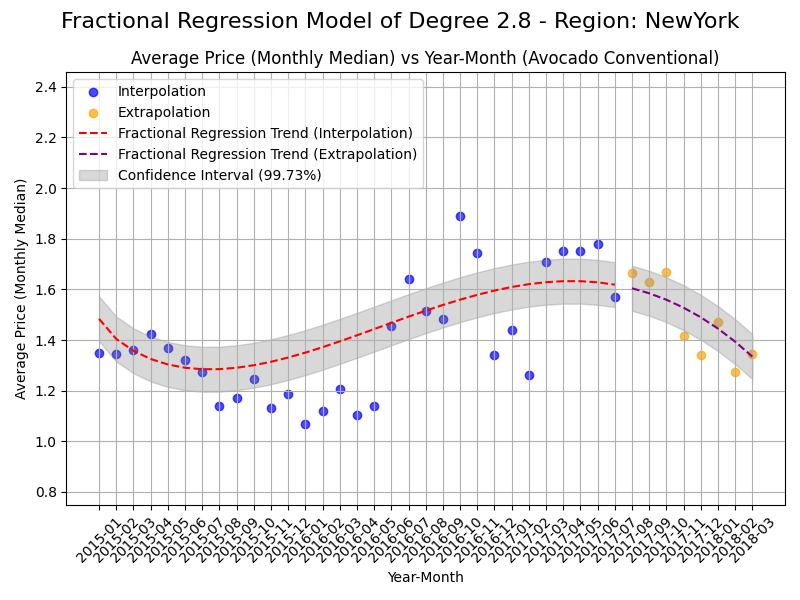}%
    }

    \caption{Comparison of metrics and fractional regression models in the NewYork region}
    \label{fig:newyork_analysis}
\end{figure}

\begin{figure}[!ht]
    \centering
    \subfloat[Box plots of average prices grouped by month]{%
        \includegraphics[width=0.48\textwidth]{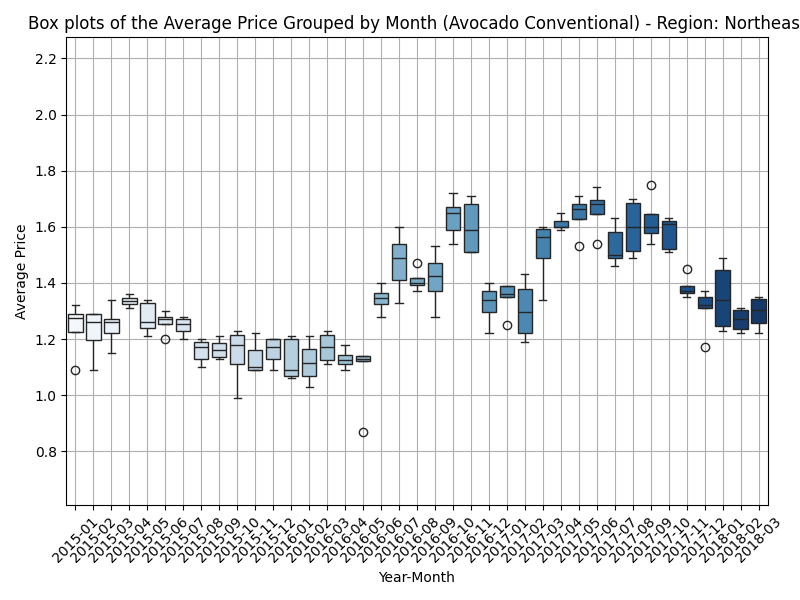}%
    }
    \hfill
    \subfloat[Logarithmic scale graph of positive $R^2$ values for interpolation and extrapolation parts simultaneously]{%
        \includegraphics[width=0.48\textwidth]{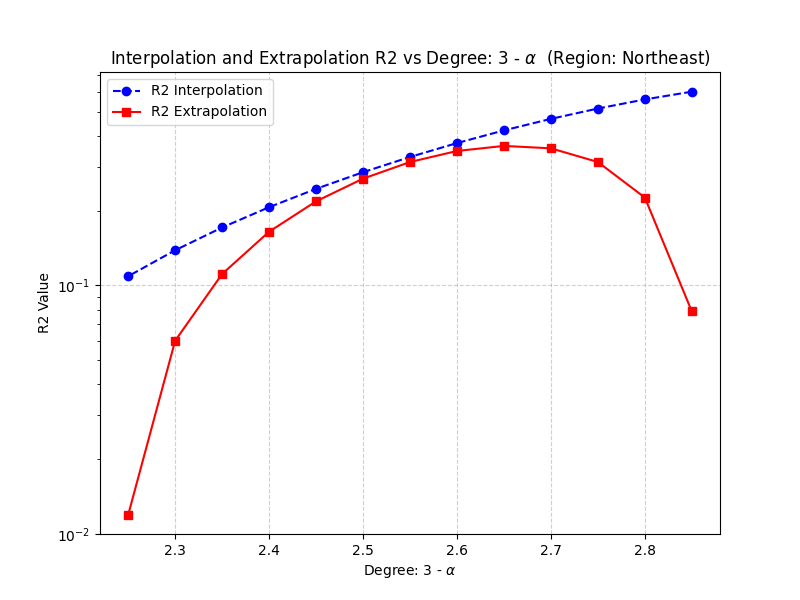}%
    }
    
    \vspace{0.1cm} 

    \subfloat[Average price distribution using the monthly median - Fractional regression model with $\alpha = 0$]{%
        \includegraphics[width=0.48\textwidth]{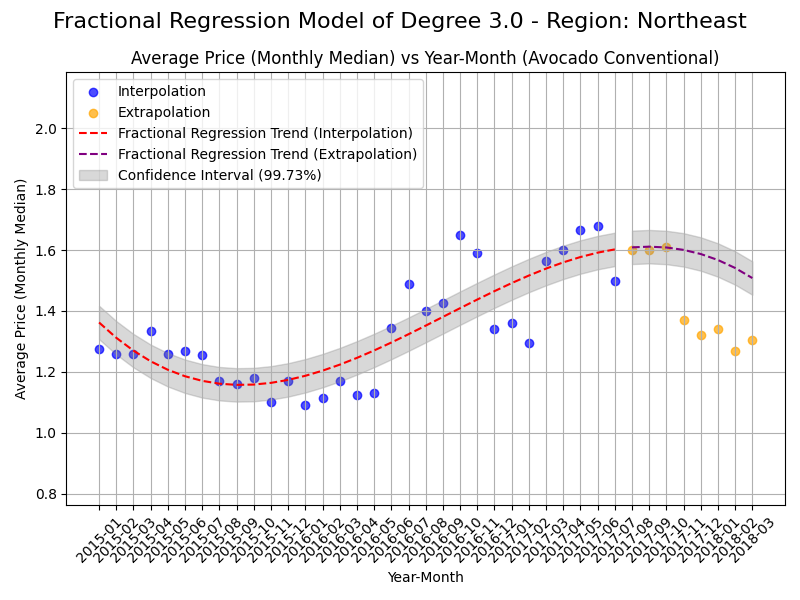}%
    }
    \hfill
    \subfloat[Average price distribution using the monthly median - Fractional regression model with $\alpha \neq 0$]{%
        \includegraphics[width=0.48\textwidth]{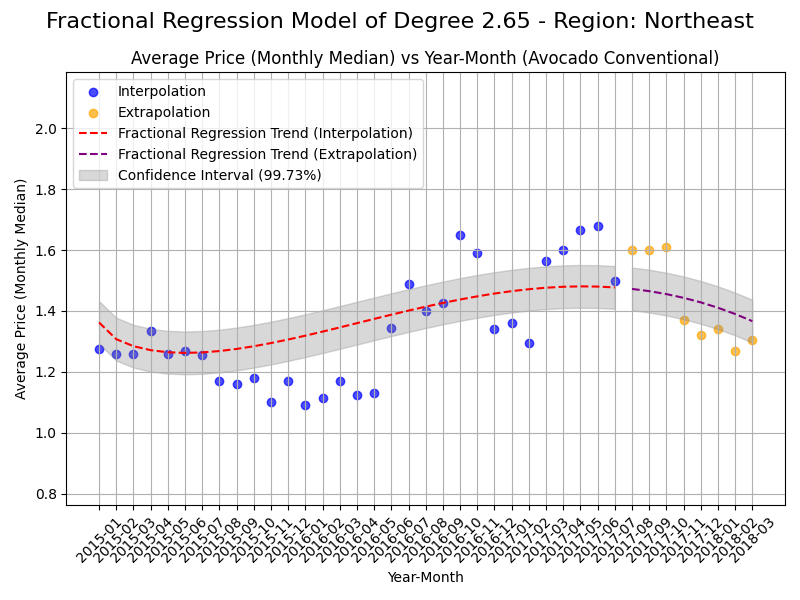}%
    }

    \caption{Comparison of metrics and fractional regression models in the Northeast region }
    \label{fig:northeast_analysis}
\end{figure}

\begin{figure}[!ht]
    \centering
    \subfloat[Box plots of average prices grouped by month]{%
        \includegraphics[width=0.48\textwidth]{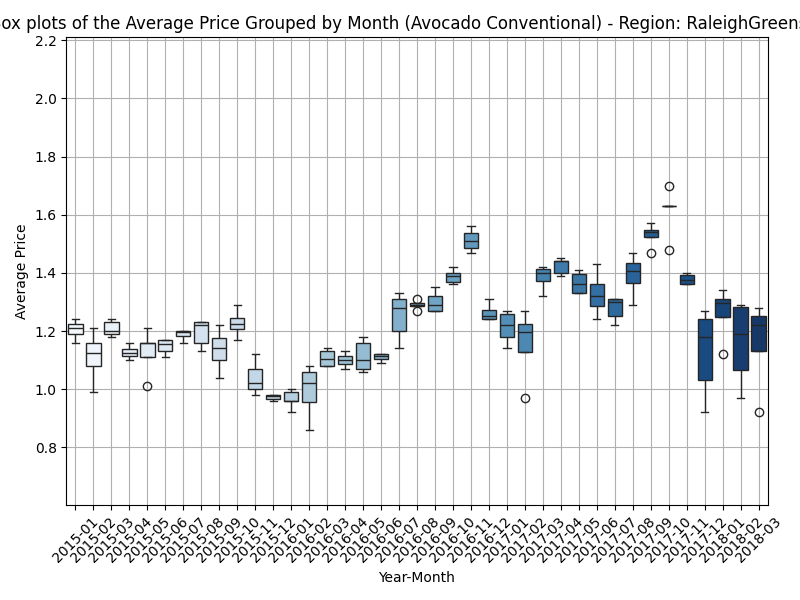}%
    }
    \hfill
    \subfloat[Logarithmic scale graph of positive $R^2$ values for interpolation and extrapolation parts simultaneously]{%
        \includegraphics[width=0.48\textwidth]{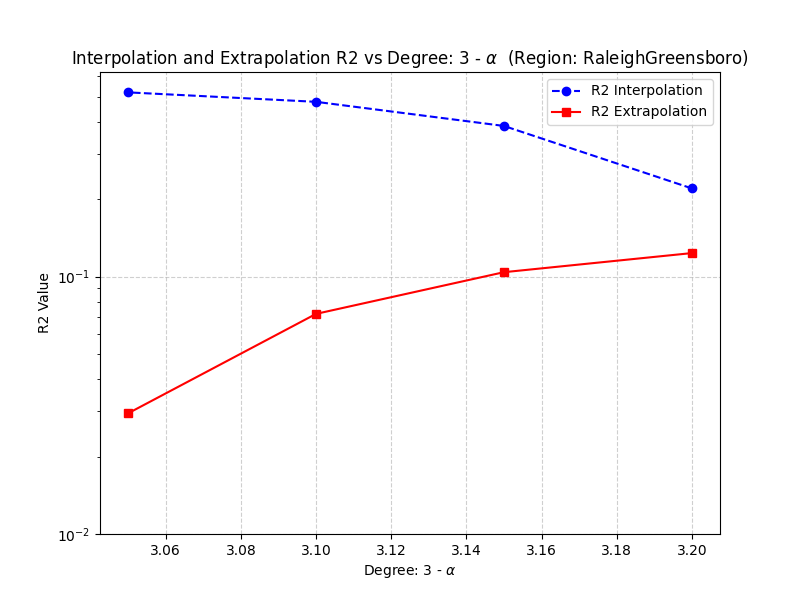}%
    }
    
    \vspace{0.1cm} 

    \subfloat[Average price distribution using the monthly median - Fractional regression model with $\alpha = 0$]{%
        \includegraphics[width=0.48\textwidth]{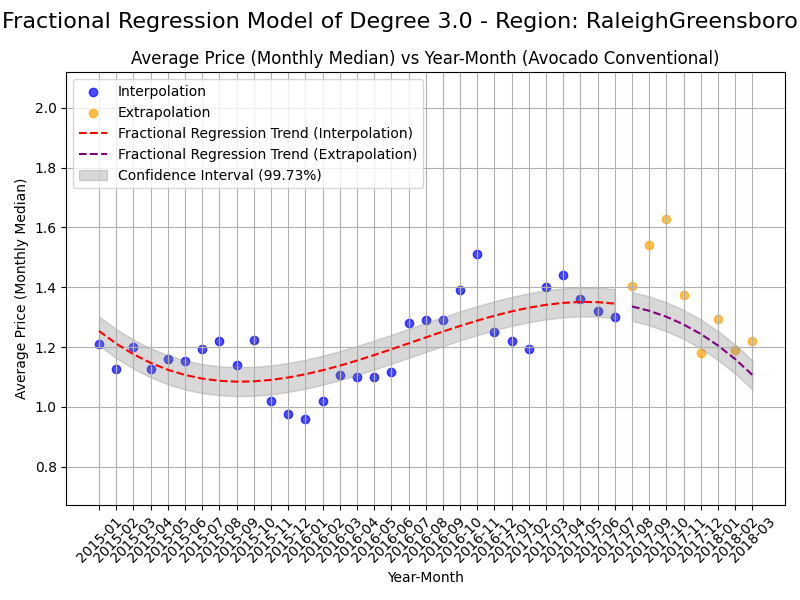}%
    }
    \hfill
    \subfloat[Average price distribution using the monthly median - Fractional regression model with $\alpha \neq 0$]{%
        \includegraphics[width=0.48\textwidth]{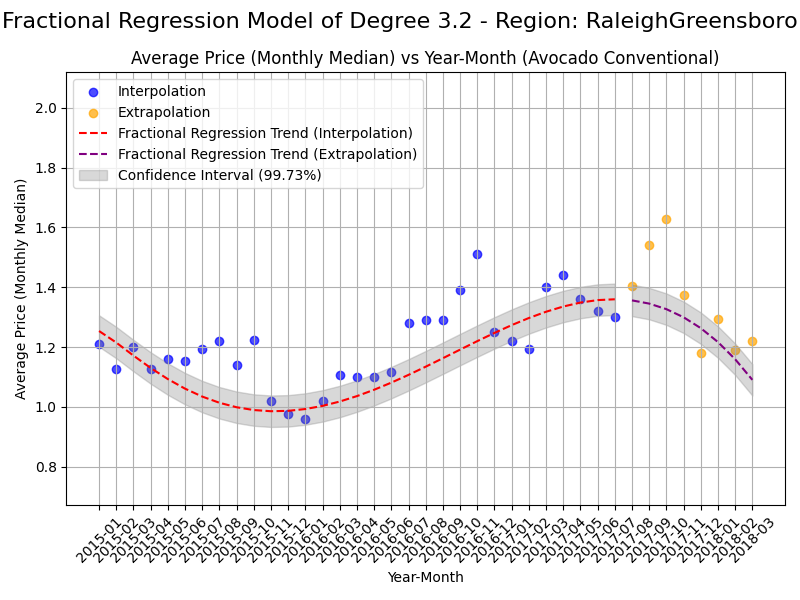}%
    }

    \caption{Comparison of metrics and fractional regression models in the RaleighGreensboro region }
    \label{fig:raleighgreensboro_analysis}
\end{figure}

\begin{figure}[!ht]
    \centering
    \subfloat[Box plots of average prices grouped by month]{%
        \includegraphics[width=0.48\textwidth]{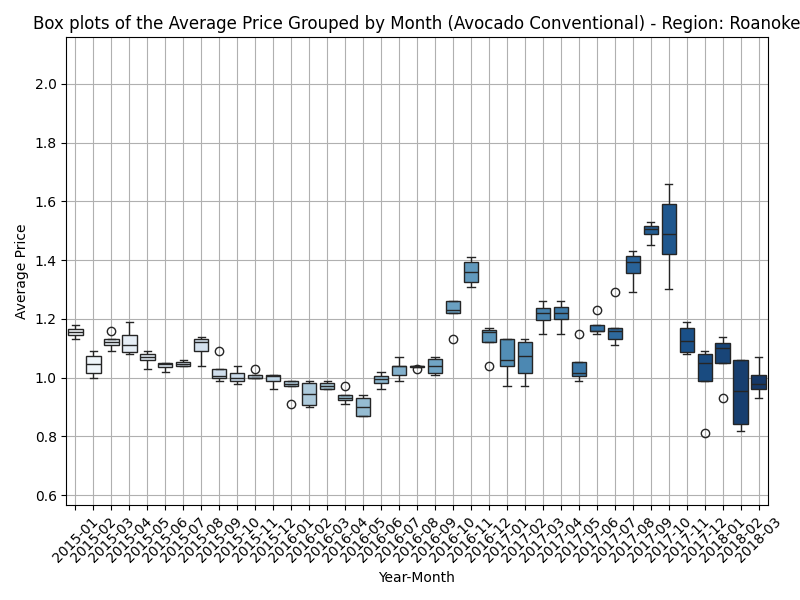}%
    }
    \hfill
    \subfloat[Logarithmic scale graph of positive $R^2$ values for interpolation and extrapolation parts simultaneously]{%
        \includegraphics[width=0.48\textwidth]{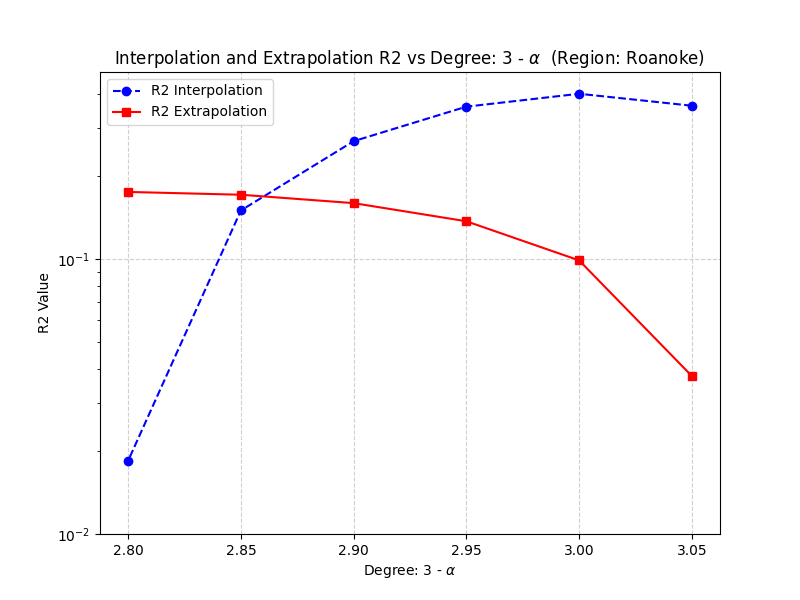}%
    }
    
    \vspace{0.1cm} 

    \subfloat[Average price distribution using the monthly median - Fractional regression model with $\alpha = 0$]{%
        \includegraphics[width=0.48\textwidth]{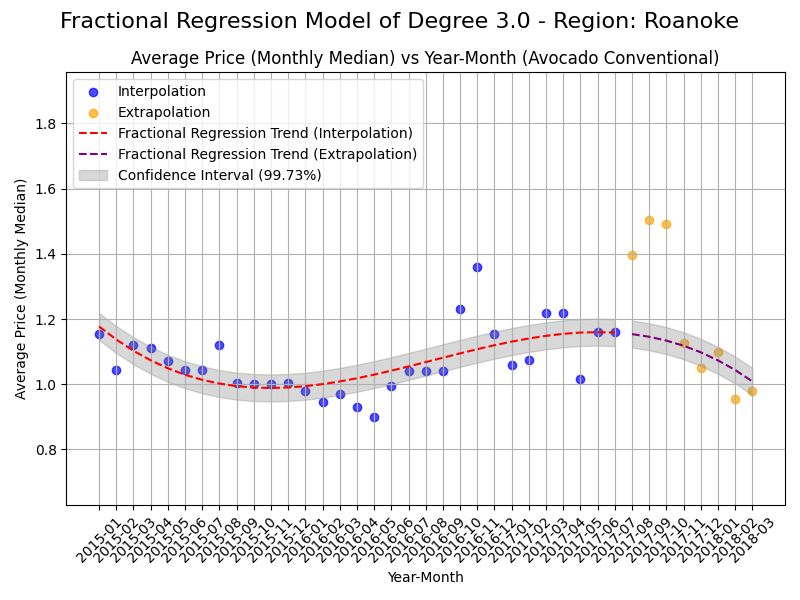}%
    }
    \hfill
    \subfloat[Average price distribution using the monthly median - Fractional regression model with $\alpha \neq 0$]{%
        \includegraphics[width=0.48\textwidth]{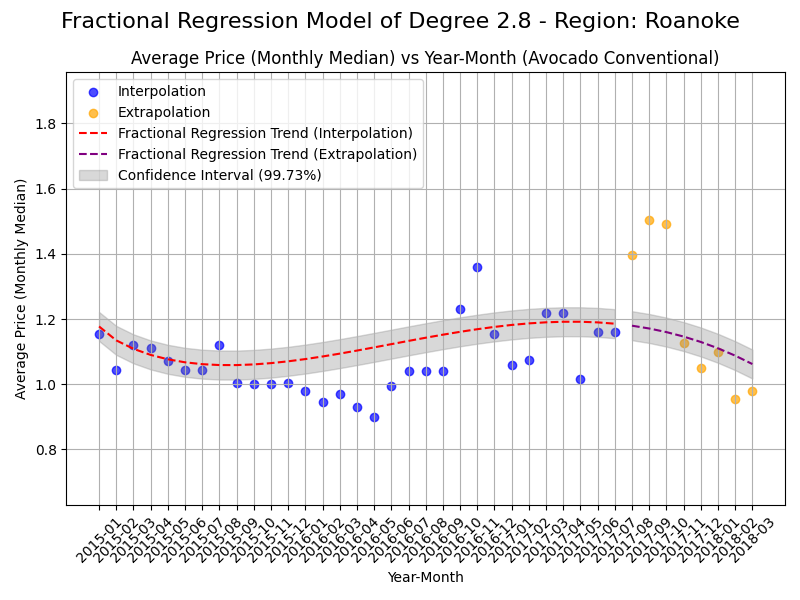}%
    }

    \caption{Comparison of metrics and fractional regression models in the Roanoke region}
    \label{fig:roanoke_analysis}
\end{figure}

\begin{figure}[!ht]
    \centering
    \subfloat[Box plots of average prices grouped by month]{%
        \includegraphics[width=0.48\textwidth]{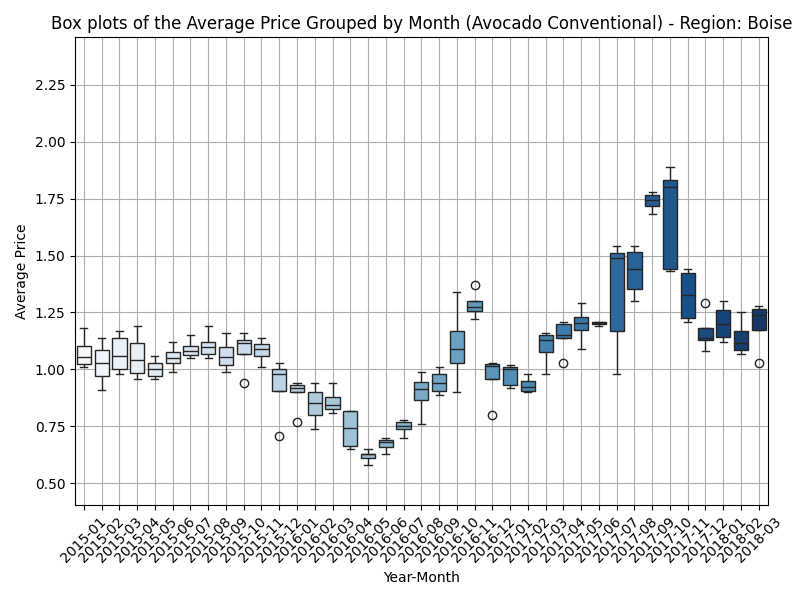}%
    }
    \hfill
    \subfloat[Logarithmic scale graph of positive $R^2$ values for interpolation and extrapolation parts simultaneously]{%
        \includegraphics[width=0.48\textwidth]{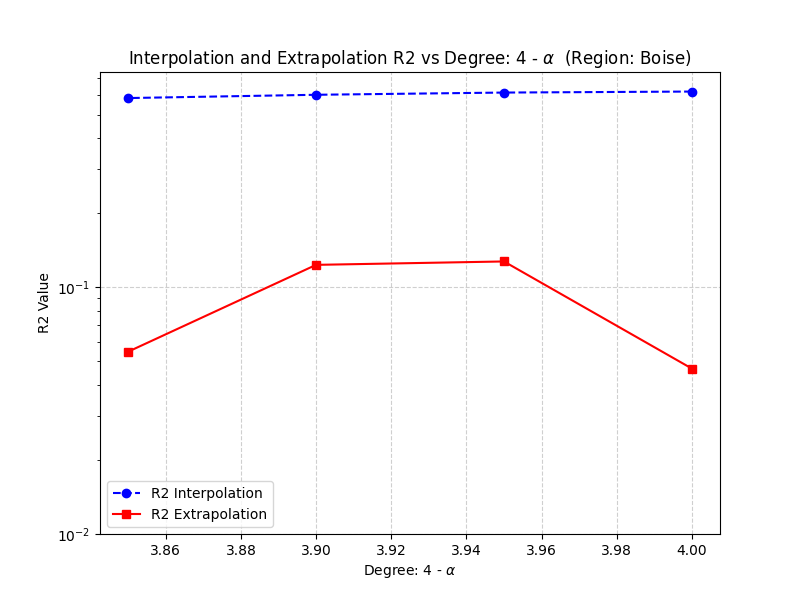}%
    }
    
    \vspace{0.1cm} 

    \subfloat[Average price distribution using the monthly median - Fractional regression model with $\alpha = 0$]{%
        \includegraphics[width=0.48\textwidth]{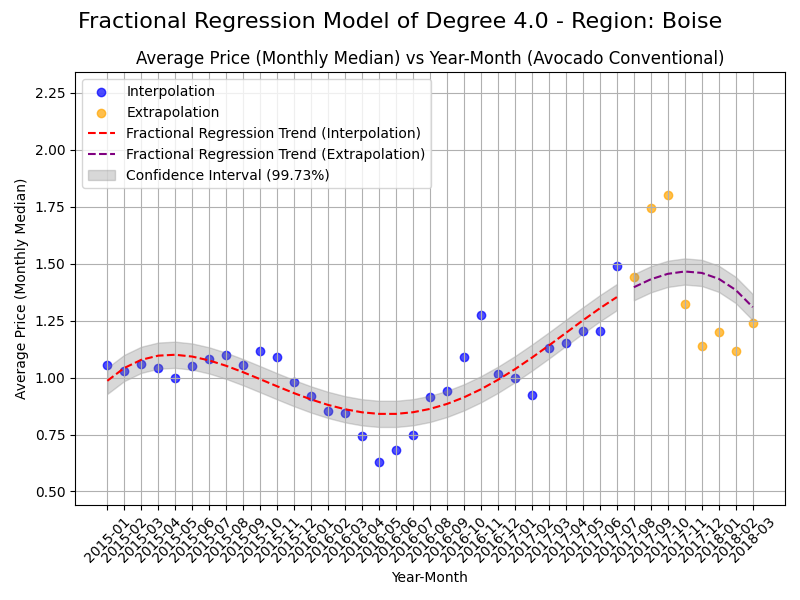}%
    }
    \hfill
    \subfloat[Average price distribution using the monthly median - Fractional regression model with $\alpha \neq 0$]{%
        \includegraphics[width=0.48\textwidth]{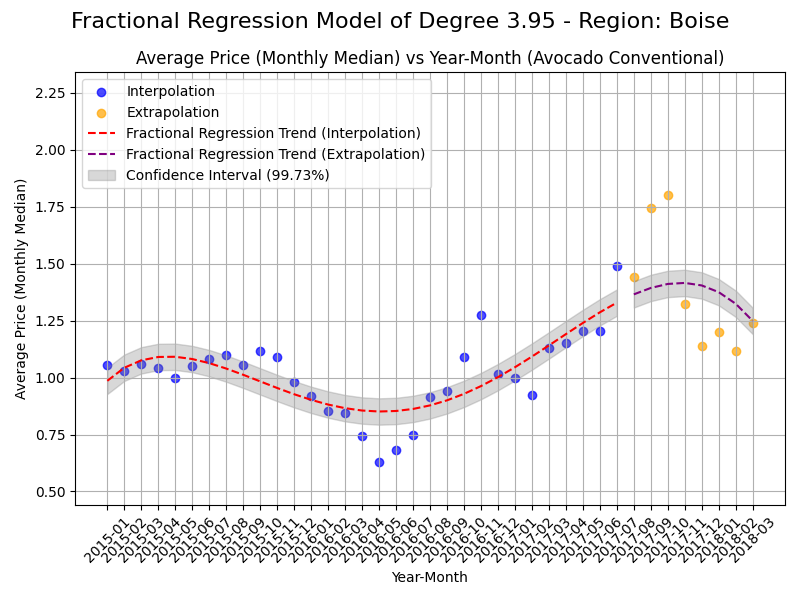}%
    }

    \caption{Comparison of metrics and fractional regression models in the Boise region }
    \label{fig:boise_analysis}
\end{figure}

\begin{figure}[!ht]
    \centering
    \subfloat[Box plots of average prices grouped by month]{%
        \includegraphics[width=0.48\textwidth]{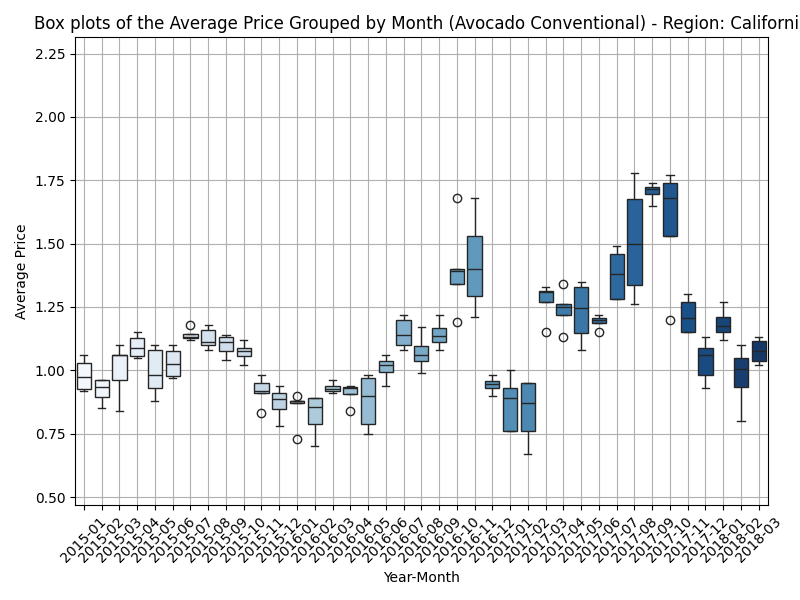}%
    }
    \hfill
    \subfloat[Logarithmic scale graph of positive $R^2$ values for interpolation and extrapolation parts simultaneously]{%
        \includegraphics[width=0.48\textwidth]{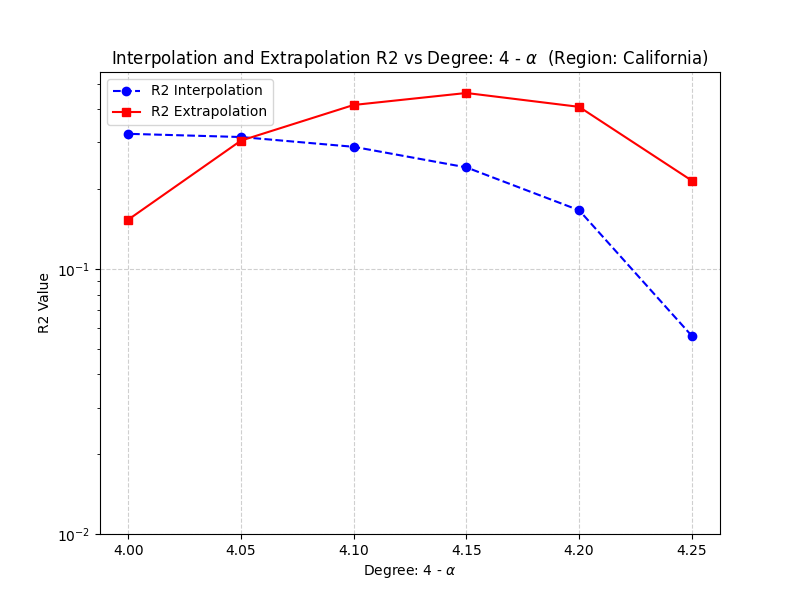}%
    }
    
    \vspace{0.1cm} 

    \subfloat[Average price distribution using the monthly median - Fractional regression model with $\alpha = 0$]{%
        \includegraphics[width=0.48\textwidth]{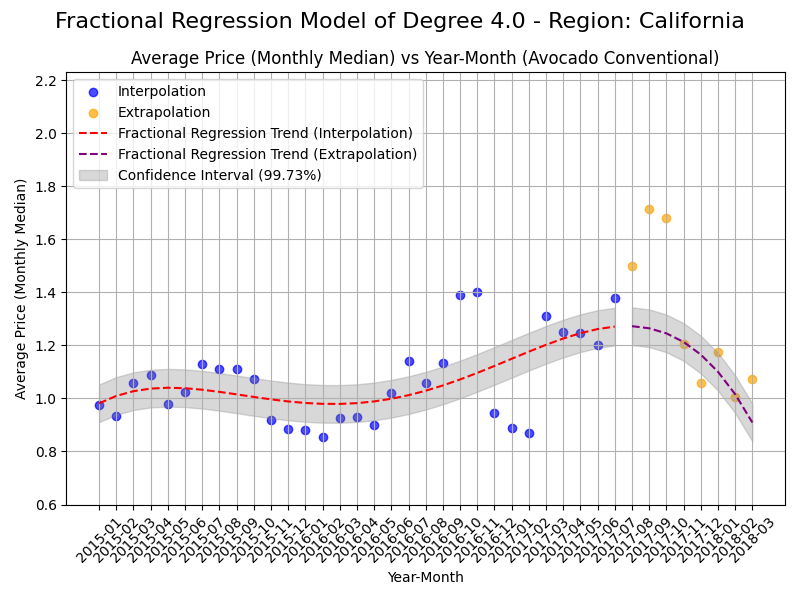}%
    }
    \hfill
    \subfloat[Average price distribution using the monthly median - Fractional regression model with $\alpha \neq 0$]{%
        \includegraphics[width=0.48\textwidth]{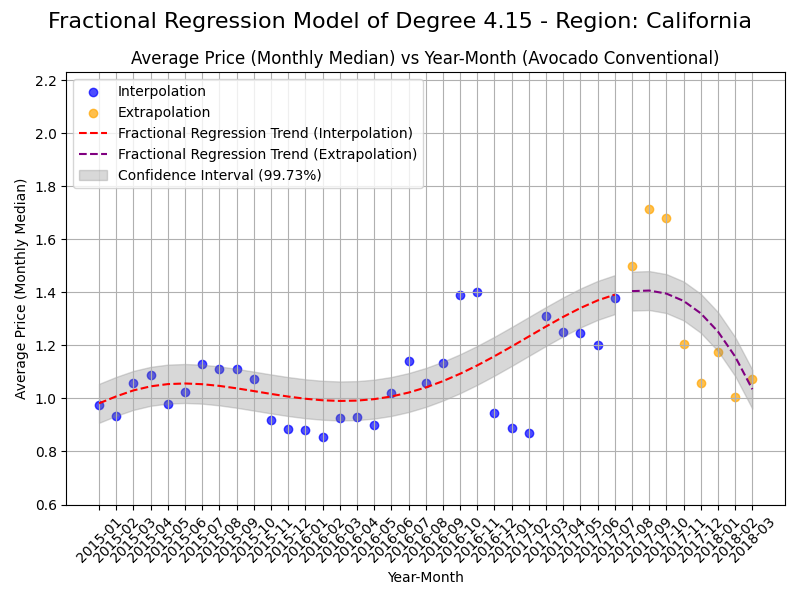}%
    }

    \caption{Comparison of metrics and fractional regression models in the California region }
    \label{fig:california_analysis}
\end{figure}

\begin{figure}[!ht]
    \centering
    \subfloat[Box plots of average prices grouped by month]{%
        \includegraphics[width=0.48\textwidth]{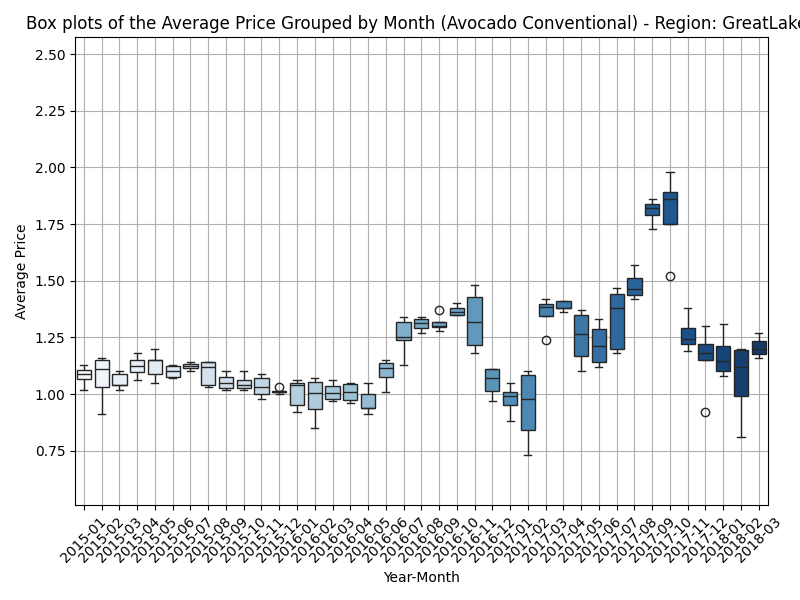}%
    }
    \hfill
    \subfloat[Logarithmic scale graph of positive $R^2$ values for interpolation and extrapolation parts simultaneously]{%
        \includegraphics[width=0.48\textwidth]{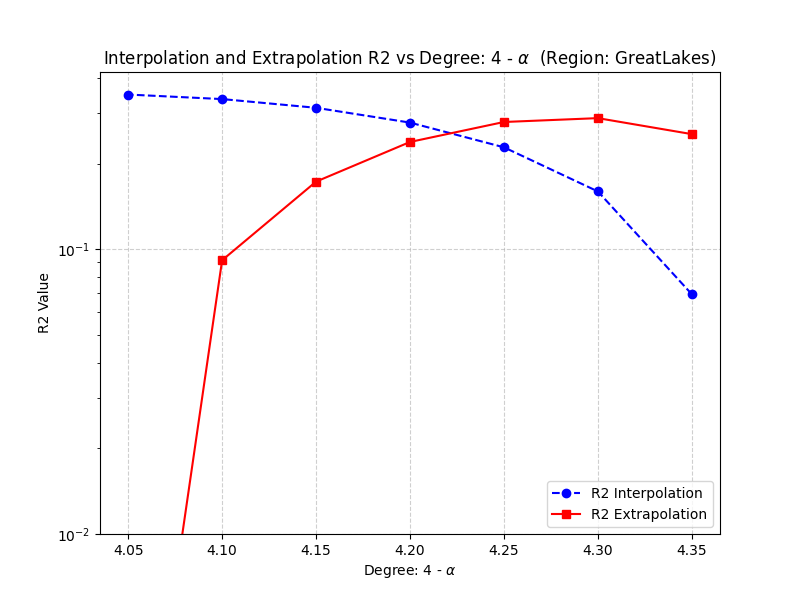}%
    }
    
    \vspace{0.1cm} 

    \subfloat[Average price distribution using the monthly median - Fractional regression model with $\alpha = 0$]{%
        \includegraphics[width=0.48\textwidth]{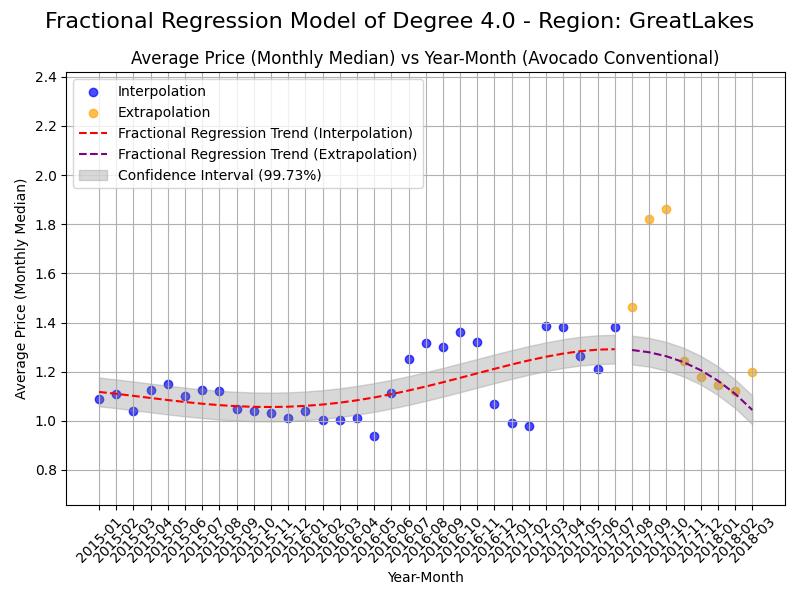}%
    }
    \hfill
    \subfloat[Average price distribution using the monthly median - Fractional regression model with $\alpha \neq 0$]{%
        \includegraphics[width=0.48\textwidth]{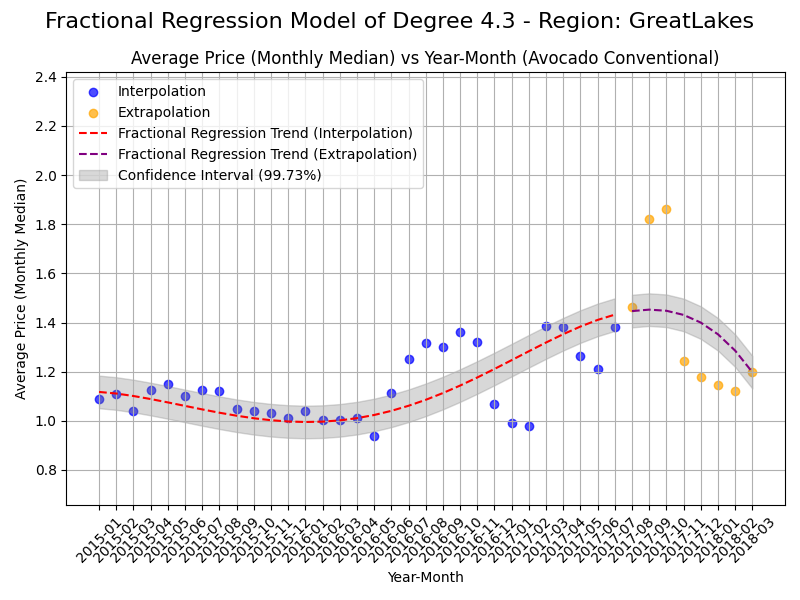}%
    }

    \caption{Comparison of metrics and fractional regression models in the GreatLakes region }
    \label{fig:greatlakes_analysis}
\end{figure}

\begin{figure}[!ht]
    \centering
    \subfloat[Box plots of average prices grouped by month]{%
        \includegraphics[width=0.48\textwidth]{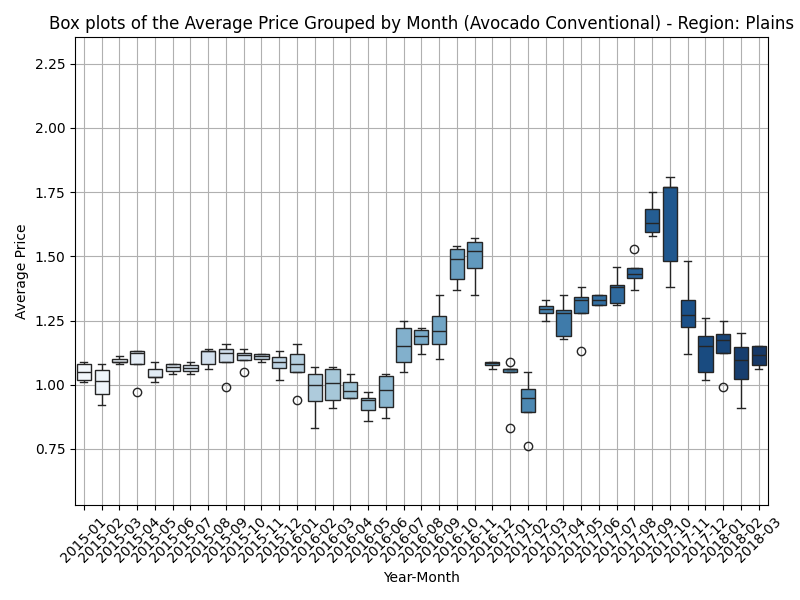}%
    }
    \hfill
    \subfloat[Logarithmic scale graph of positive $R^2$ values for interpolation and extrapolation parts simultaneously]{%
        \includegraphics[width=0.48\textwidth]{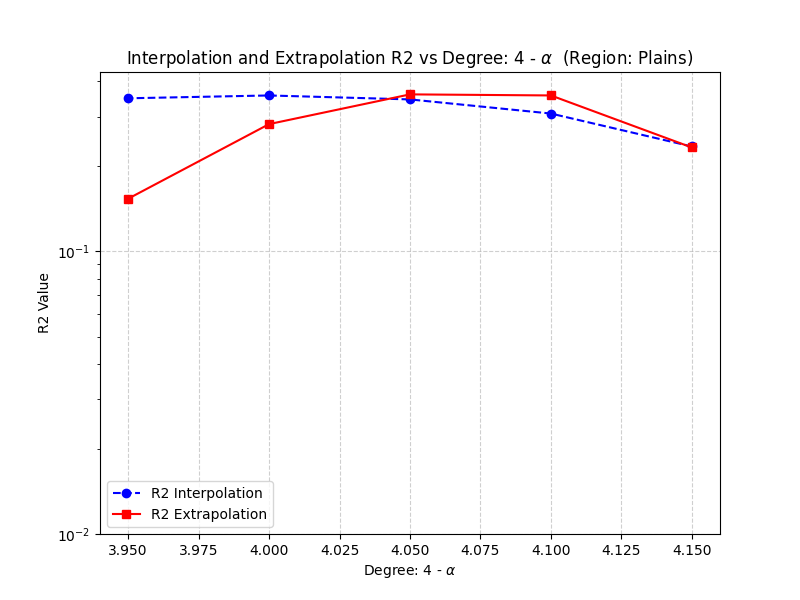}%
    }
    
    \vspace{0.1cm} 

    \subfloat[Average price distribution using the monthly median - Fractional regression model with $\alpha = 0$]{%
        \includegraphics[width=0.48\textwidth]{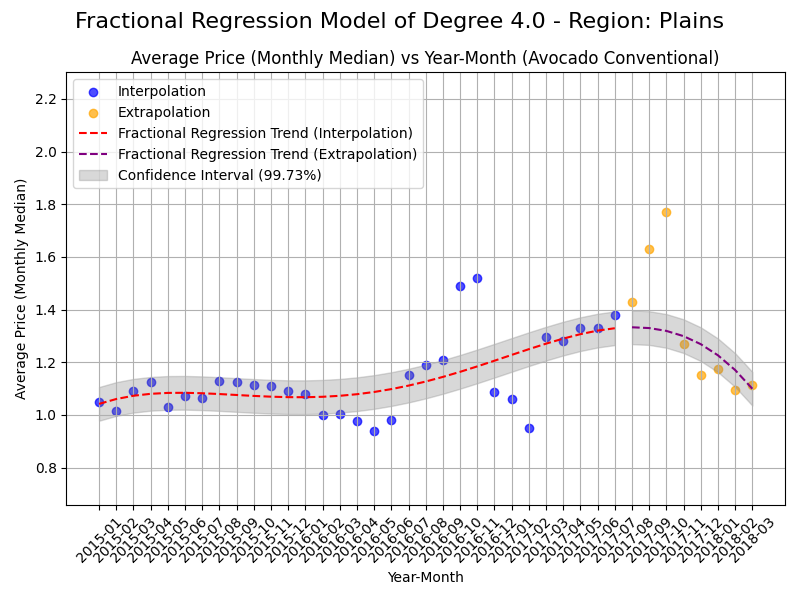}%
    }
    \hfill
    \subfloat[Average price distribution using the monthly median - Fractional regression model with $\alpha \neq 0$]{%
        \includegraphics[width=0.48\textwidth]{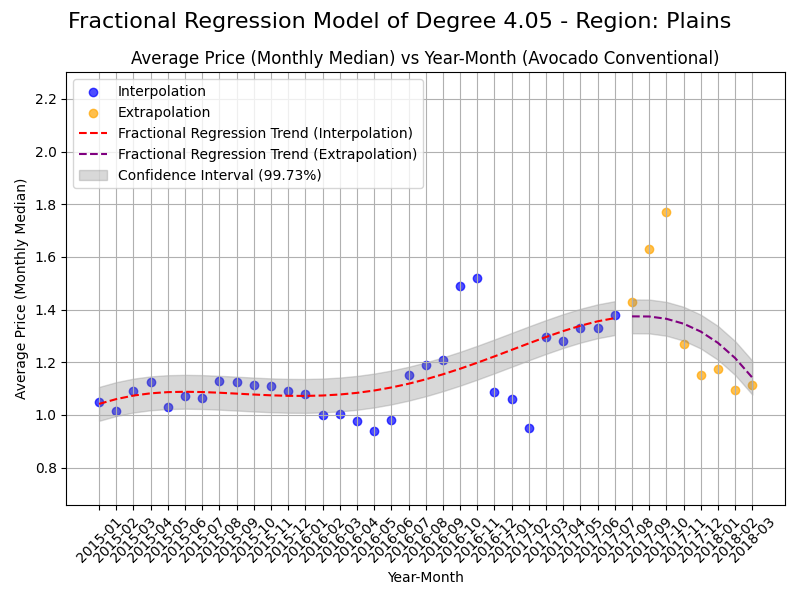}%
    }

    \caption{Comparison of metrics and fractional regression models in the Plains region}
    \label{fig:plains_analysis}
\end{figure}

\begin{figure}[!ht]
    \centering
    \subfloat[Box plots of average prices grouped by month]{%
        \includegraphics[width=0.48\textwidth]{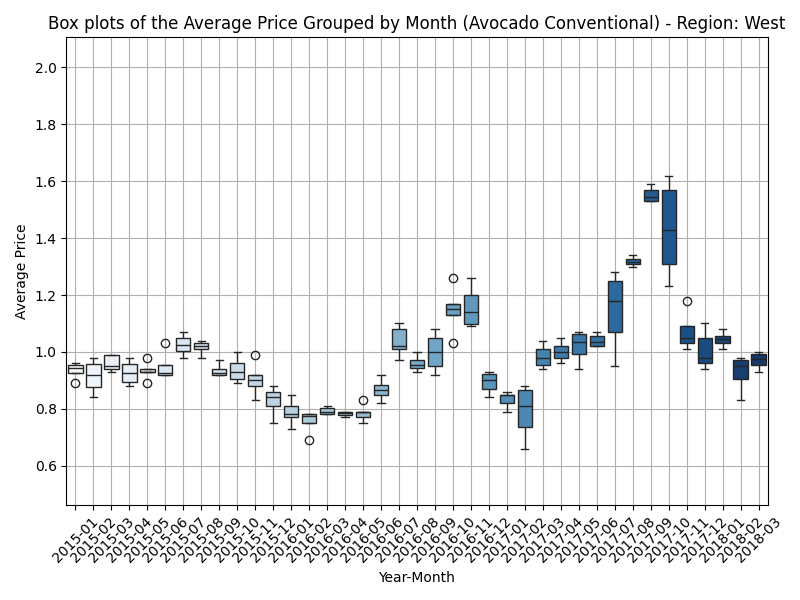}%
    }
    \hfill
    \subfloat[Logarithmic scale graph of positive $R^2$ values for interpolation and extrapolation parts simultaneously]{%
        \includegraphics[width=0.48\textwidth]{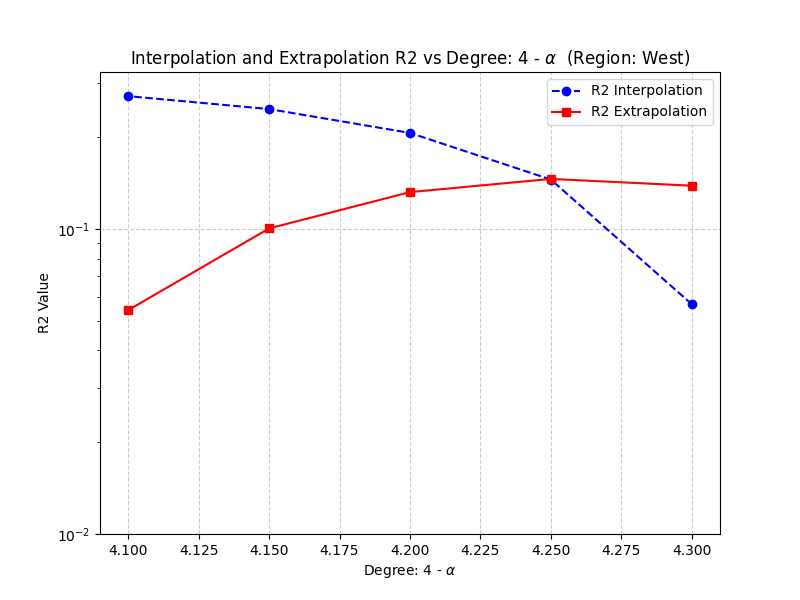}%
    }
    
    \vspace{0.1cm} 

    \subfloat[Average price distribution using the monthly median - Fractional regression model with $\alpha = 0$]{%
        \includegraphics[width=0.48\textwidth]{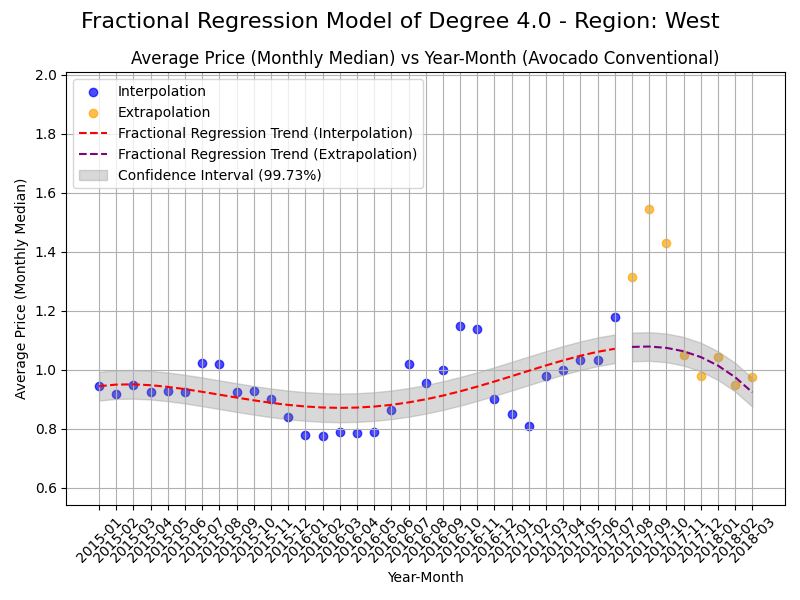}%
    }
    \hfill
    \subfloat[Average price distribution using the monthly median - Fractional regression model with $\alpha \neq 0$]{%
        \includegraphics[width=0.48\textwidth]{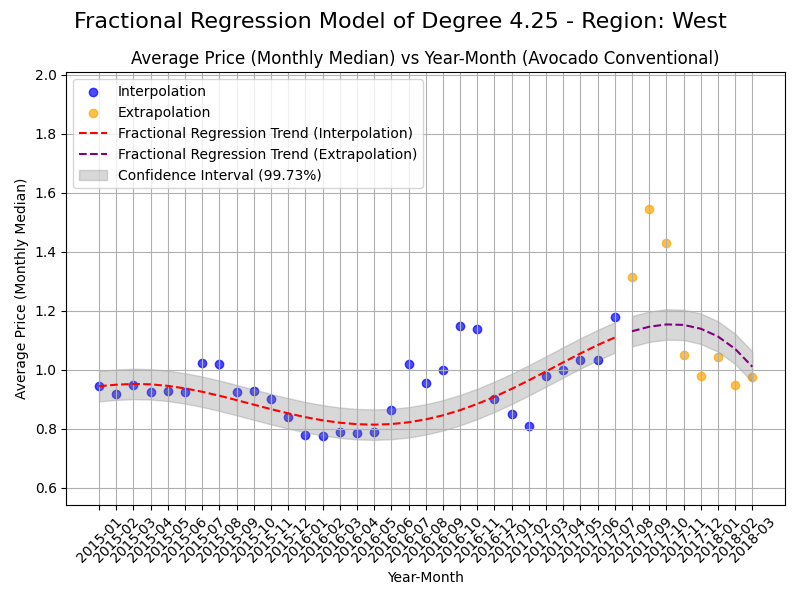}%
    }

    \caption{Comparison of metrics and fractional regression models in the West region}
    \label{fig:west_analysis}
\end{figure}

\clearpage
\newpage

\bibliographystyle{unsrt}

\bibliography{Biblio}

\end{document}